\documentclass[communication letter]{IEEEtran}
\usepackage[LGR,T1]{fontenc}
\usepackage[latin9]{inputenc}
\usepackage{color}
\usepackage{float}
\usepackage{units}
\usepackage{amsmath}
\usepackage{amssymb}
\usepackage{graphicx}
\usepackage[bookmarks=true,bookmarksnumbered=true,bookmarksopen=true,bookmarksopenlevel=1,
 breaklinks=false,pdfborder={0 0 0},pdfborderstyle={},backref=false,colorlinks=false]
 {hyperref}
\hypersetup{pdftitle={Your Title},
 pdfauthor={Your Name},
 pdfpagelayout=OneColumn, pdfnewwindow=true, pdfstartview=XYZ, plainpages=false}

\makeatletter

\DeclareRobustCommand{\greektext}{%
  \fontencoding{LGR}\selectfont\def\encodingdefault{LGR}}
\DeclareRobustCommand{\textgreek}[1]{\leavevmode{\greektext #1}}

\newcommand{\lyxmathsym}[1]{\ifmmode\begingroup\def\b@ld{bold}
  \text{\ifx\math@version\b@ld\bfseries\fi#1}\endgroup\else#1\fi}

\providecommand{\tabularnewline}{\\}
\floatstyle{ruled}
\newfloat{algorithm}{tbp}{loa}
\providecommand{\algorithmname}{Algorithm}
\floatname{algorithm}{\protect\algorithmname}

\usepackage[caption=false, font=footnotesize]{subfig}

\usepackage{algorithmic}
\usepackage{cite}

\ifdefined\showcaptionsetup
 \PassOptionsToPackage{caption=false}{subfig}
\fi
\usepackage{subfig}
\makeatother

\begin{document}
\title{Variable-Length Wideband CSI Feedback via Loewner Interpolation and
Deep Learning}
\author{Meilin Li, Wei Xu, Zhixiang Hu and An Liu,\ \textit{Senior Member},\ \textit{IEEE}\vspace{-0.7cm}
}
\maketitle
\begin{abstract}
\textcolor{black}{In this paper, we propose a variable-length wideband
channel state information (CSI) feedback scheme for Frequency Division
Duplex (FDD) massive multiple-input multiple-output (MIMO) systems
in U6G band (6425MHz-7125MHz). }Existing compressive sensing (CS)-based
and deep learning (DL)-based schemes preprocess the channel by truncating
it in the angular-delay domain. However, the energy leakage effect
caused by the Discrete Fourier Transform (DFT) basis will be more
serious and leads to a bottleneck in recovery accuracy when applied
to wideband channels such as those in U6G.\textcolor{black}{{} To solve
this problem, we introduce the Loewner Interpolation (LI) framework
which generates a set of dynamic bases based on the current CSI matrix,
enabling highly efficient compression in the frequency domain. Then,
the LI basis is further compressed in the spatial domain through a
neural network. To achieve a flexible trade-off between feedback overhead
and recovery accuracy, we design a rateless auto-encoder trained with
tail dropout and a multi-objective learning schedule, supporting variable-length
feedback with a singular model. Meanwhile, the codewords are ranked
by importance, ensuring that the }base station (BS)\textcolor{black}{{}
can still maintain acceptable reconstruction performance under limited
feedback with tail erasures. Furthermore, an adaptive quantization
strategy is developed for the feedback framework to enhance robustness.
}Simulation results demonstrate that the proposed scheme could achieve
higher CSI feedback accuracy with less or equal feedback overhead,
and improve spectral efficiency compared with baseline schemes.
\end{abstract}


\section{Introduction}

Massive multiple-input multiple-output (MIMO) technology is a key
technology for enhancing system capacity and spectral efficiency in
5G and beyond systems\ \cite{MIMO_1}\cite{MIMO_2}. For Frequency
Division Duplex (FDD) systems, the reciprocity between the uplink
channel and the downlink channel is weak. The base station (BS) obtains
accurate downlink channel state information (CSI) by sending downlink
pilots and receiving feedback from the user equipment (UE). In 2022,
the 3rd Generation Partnership Project (3GPP) officially authorized
the U6G spectrum (6425MHz\textasciitilde 7125MHz), marking this band
as a potential spectrum hotspot in the next generation communication
systems. With such huge bandwidth and the increasing number of transmit
antennas, the dimension of the CSI matrix significantly increases
and leads to a large feedback overhead\ \cite{Overhead_1}\cite{Overhead_2}.
To leverage the advantages of massive MIMO systems in U6G band, ensuring
high-precision CSI feedback while controlling the feedback overhead
becomes a key technical challenge.

\subsection{Related Works}

For CSI feedback in FDD massive MIMO systems, a wide range of approaches
have been proposed. Generally, these methods can be classified into
three categories: codebook-based feedback schemes, compressed sensing
(CS)-based feedback schemes and deep learning (DL)-based feedback
schemes.

For codebook-based CSI feedback, some works design codebooks by exploiting
channel sparsity in the angular and/or delay domains\ \cite{code_book_1,3gpp214r16,code_book_2}.
For instance, \cite{code_book_1}\ proposed a subspace codebook that
adapts to the angle of departure (AoD). However, it relies on the
assumption that the AoD support remains stable over a relatively long
coherence interval. 3GPP Release-16 (R16) proposed the Enhanced Type
II (eType II) codebook\ \cite{3gpp214r16}, which leverages the sparsity
in the angular and delay domains via a Discrete Fourier Transform
(DFT) basis.\textcolor{red}{{} }However, the eType II codebook in R16
is constrained by both the number of antenna ports (maximum 32) and
the supported bandwidth. For massive MIMO systems envisioned for U6G,
these limitations, together with the energy leakage inherent to the
fixed DFT basis, lead to notable deterioration in CSI reconstruction
accuracy. Some subsequent works have focused on designing precoders
to increase channel sparsity. \cite{code_book_2}\ proposed an eigenvector-based
precoder that leverages partial angle-delay reciprocity between uplink
and downlink channels. This approach, however, requires precise channel
statistics, which are often difficult to obtain in practice.

For CS-based CSI feedback, the inherent correlation structure within
the channel matrix enables a compact and sparse representation in
a suitable transform domain. CS theory thus provides an efficient
mechanism for CSI feedback by formulating it as a sparse signal recovery
problem. Conventional\textcolor{red}{{} }CS algorithms like Least Absolute
Shrinkage and Selection Operator (LASSO) and Approximate Message Passing
(AMP) are constrained by their reliance on simple sparsity assumptions.
To overcome these limitations, more advanced CS-based approaches,
including BM3D-AMP\ \cite{CS_4} and TVAL3\ \cite{CS_5}, have been
developed to exploit richer prior models.\textcolor{red}{{} }Also, some
studies further exploit correlations of the channels across different
domains to design CS-based CSI feedback schemes. For example, \cite{CS_1}\textcolor{blue}{\ }proposed
a hybrid adaptive feedback scheme leveraging the optimality of the
Karhunen-Loeve Transform (KLT) and the sparsity in the spatial and
frequency domains arising from spatially correlated antenna arrays.
In\ \cite{CS_2}, a partial channel support information (P-CSPI)-aided
burst LASSO algorithm was developed to utilize the temporal correlation
of the channel support, thereby reducing pilot and feedback overhead.
Although CS-based feedback methods have a solid theoretical foundation,
most of them rely on iterative reconstruction which leads to high
computational complexity and struggle to achieve higher recovery accuracy.

3GPP R18/19 further introduces artificial intelligence (AI) to CSI
feedback as a promising key technology. DL-based CSI feedback schemes
treat downlink CSI as an image and aim to reconstruct the original
image from its latent space representation. Many studies\ \cite{DL_1,DL_2,DL_3,DL_4}
have achieved high reconstruction accuracy by refining the architecture
of the auto-encoder. However, these methods typically rely on DFT
and truncation as preprocessing steps of the original channel matrix.
In wideband scenarios, the energy leakage problem caused by DFT is
more serious. A large truncation ratio results in significant precision
loss before network compression, forming a bottleneck in reconstruction
accuracy, while a small truncation ratio significantly increases the
computational complexity of the network. Furthermore, most of these
approaches neglect the reconstruction error caused by codeword quantization
and are limited to a single compression ratio (CR).

Several recent studies have attempted to overcome the aforementioned
limitations by removing the dependency on DFT-based preprocessing
and enabling multi-rate CSI feedback with\textcolor{blue}{{} }codeword
quantization\textcolor{blue}{. \cite{DL_8}}\ proposed a compressive
sampled CSI feedback method which combines sampling with neural networks.
Instead of using the DFT basis, the compression in the frequency domain
is achieved through a sampling network and an interpolation network.
However, the application scenario for\textcolor{blue}{\ \cite{DL_8}}
is limited to narrowband scenarios, where the number of subcarriers
used in simulation is relatively small. When extended to wideband
channels with a significantly larger number of subcarriers, the computational
complexity of the network increases sharply due to the higher dimension
of the input tensor, potentially leading to increased training cost,
longer inference time, and reduced practicality in U6G massive MIMO
systems. In\textcolor{blue}{\ \cite{DL_5}}, multi-rate feedback
with quantization is considered based on the proposed serial multiple-rate
framework and parallel multiple-rate framework. However, the BS needs
to deploy several pre-trained decoders according to the corresponding
CRs and the length of the feedback codewords is still fixed. \cite{DL_6}
proposed an integrated, bit-level feedback mechanism with variable-length
feedback adaptability which takes the current CSI matrix directly
as the input of the auto-encoder and arranges a large number of self-attention
mechanisms \ \cite{DL_7} at the BS. Simulation results indicate
that though a flexible trade-off between feedback overhead and accuracy
has been achieved, the recovery accuracy of the CSI matrix is low
and the computational complexity is higher than that of other convolutional
neural networks (CNNs).

Although there is a large body of work on CSI feedback, most of these
methods have one of the following drawbacks. First, they focus more
on the CR and ignore the reconstruction error introduced by the quantization
of the feedback codewords. Second, the CRs supported by these schemes
are limited, and the length of the output codewords is still fixed
which poses challenges for practical deployment in some scenarios
such as \textit{channels with tail erasures}\ \footnote{A channel with tail erasures is suitable for the situation where the
receiver effectively stops obtaining meaningful symbols after an unknown
point in time. In this scenarios, only an initial portion of the codeword
is reliably received, either due to abrupt link interruption or due
to a system intentionally decoding\ \cite{channel_with_tail_erasures}.}. Third, they preprocess the channel matrix with DFT and consider
the reconstruction of the truncated sparse angular-delay channel $\tilde{\mathbf{H}}=f_{trun}(\mathbf{F}_{d}\mathbf{H}\mathbf{F}_{a}^{H})$,
where $f_{trun}(\cdot)$ denotes the truncation operator. Although
some existing methods can achieve high-precision recovery of $\tilde{\mathbf{H}}$,
the energy leakage problem caused by DFT and truncation in the wideband
scenario is more serious, resulting in a reconstruction accuracy bottleneck
induced by $f_{trun}(\cdot)$. In summary, achieving high-accuracy
CSI feedback and a flexible trade-off between overhead and precision
with quantization remains a key issue in massive MIMO systems applied
to the U6G band.

\subsection{Key Contributions}

To address the issues mentioned above, this paper proposes a variable-length
wideband CSI feedback based on Loewner Interpolation (LI) framework
and DL. This paper substantially extends our prior work \ \cite{WCL_LML},
where a preliminary LI-based compression scheme was introduced. In
this paper, the framework is further generalized to the frequency
domain and integrated with a rateless DL-based spatial compression
architecture and a robust quantization strategy, providing a more
comprehensive system framework. The main contributions of this paper
are summarized as follows:
\begin{itemize}
\item LI-based compression in the frequency domain: Unlike most of the DL-based
CSI feedback schemes preprocessing the CSI matrix with DFT and truncation,
we introduce the LI framework to learn a set of effective bases from
the current CSI matrix in a straightforward manner and ensure perfect
recovery at sample points based on the system-theoretical methodology.
Then the Model Order Reduction (MOR) in the LI framework helps to
prevent the expansion of the parameters and achieve flexible compression
in the frequency domain. As a result, we obtain a set of LI basis
matrices $\left\{ \mathbf{A\mathrm{,}B\mathrm{,}C}\right\} $ with
simple structures.\textcolor{blue}{{} }In particular, $\mathbf{A}\in\mathbb{C}^{r_{f}\times r_{f}}$
is a diagonal matrix with $r_{f}$ denoting the order preserved in
the frequency domain, and $\mathbf{B}\in\mathbb{C}^{r_{f}\times2}$
is a low-dimensional matrix with the dimension $2$ arising from the
dual-polarized antenna configuration. Therefore, $\mathbf{A}$ and
$\mathbf{B}$ only contain a small number of non-zero elements. On
the other hand, $\mathbf{C}\in\mathbb{C}^{N_{t}\times r_{f}}$ , where
$N_{t}$ represents the number of antennas, which is large for massive
MIMO. As such, the feedback overhead of $\mathbf{C}$ is dominated
by the spatial dimension of $\mathbf{C}$. 
\item Rateless auto-encoder-based compression in the spatial domain: After
obtaining the LI bases, we design a cascaded spatial compression scheme
to further reduce the feedback overhead caused by the large spatial
dimension of $\mathbf{C}$. Specifically, a matrix multiplication
operator $\mathbf{F}\left(\cdot\right)$ is first designed to transform
$\mathbf{C}$ into another matrix $\mathbf{F}\left(\mathbf{C}\right)$
to prevent the error amplification caused by algorithm cascading and
sparsity enhancement. Then, the transformed matrix $\mathbf{F}\left(\mathbf{C}\right)$
is fed into a rateless auto-encoder, which generates the corresponding
codeword, denoted by $\mathbf{v}$. Furthermore, the rateless auto-encoder
is trained with \textcolor{black}{tail dropout and a multi-objective
learning schedule} to implement variable-length feedback of the codewords.
\item Robust quantization algorithm for the overall framework: To account
for quantization effects in the feedback scheme, we design different
quantization schemes for all the codewords $\left\{ \mathbf{A},\mathbf{B},\mathbf{v}\right\} $
respectively based on their numerical structures. Although the quantization
bits of $\mathbf{v}$ dominate the overall feedback overhead, the
reconstruction error of $\mathbf{H}$ is more sensitive to the reconstruction
errors of $\left\{ \mathbf{A},\mathbf{B}\right\} $ than that of $\mathbf{C}$,
which is primarily determined by the quantization error of $\mathbf{v}$
and the CR of the auto-encoder. A slight refinement of the quantization
bit allocation for $\left\{ \mathbf{A},\mathbf{B}\right\} $ can effectively
enhance the robustness of the entire recovery process. Motivated by
this observation, we develop an adaptive bit-allocation algorithm
for $\left\{ \mathbf{A},\mathbf{B}\right\} $, achieving notable gains
with only negligible additional feedback overhead. First, the algorithm
computes the gradient matrix of the reconstruction error to identify
potential low-accuracy recovery samples. Based on this information,
the system proactively adjusts the quantization bit allocation at
the UE to prevent the potential abnormal recovery at the BS. Simulation
results demonstrate that the proposed algorithm significantly enhances
robustness with a negligible overhead increase.
\end{itemize}
The rest of the paper is organized as follows. Section\ \ref{sec:2}
presents the system model for the CSI feedback in FDD massive MIMO
systems. In Section\ \ref{sec:3}, we introduce the LI framework
for the CSI compression in the frequency domain. Then, the further
spatial-domain compression based on the rateless auto-encoder is introduced
in Section\ \ref{sec:4}. The robust quantization algorithm, along
with the summary of the overall framework, is presented in Section\ \ref{sec:5}.
In Section\ \ref{sec:6}, we provide simulation results to demonstrate
the performance of the proposed scheme. Finally, we give the conclusion
of the paper in Section\ \ref{sec:Conclusion}.

In this paper, we adopt the following notational conventions: $a$
is a scalar; $\mathbf{a}$ is a vector; $\mathbf{A}$ is a matrix.
$\left(\cdot\right)^{T},\left(\cdot\right)^{H},\left(\cdot\right)^{*},diag\left(\cdot\right)$,$\left\Vert \cdot\right\Vert _{F}$
and $Tr\left(\cdot\right)$ denote the transpose, conjugate transpose,
conjugate, diagonal matrix, Frobenius norm and trace respectively.
$\mathbb{C}^{a\times b}$ is a matrix space with $a$ rows and $b$
columns. $\left(\mathbf{A}\right)_{i,j}$ denotes the element in the
$i$-th row and the $j$-th column of the matrix $\mathbf{A}$ and
$\left(\mathbb{A}\right)_{i,j}$ denotes the submatrix located at
the $i$-th row and the $j$-th column of the block matrix $\mathbb{A}$.
$\mathbf{I}_{N}$ and $\mathbf{0}_{N}$ represent an $N\times N$
identity matrix and a $N\times N$ zero matrix.  $\triangleq$ refers
to the definition symbol.

\section{\textcolor{black}{System Model \label{sec:2}}}

We consider a FDD massive MIMO system where the BS is equipped with
$N_{t}$ antennas and each UE has $N_{r}$ antennas. The downlink
channel at a given subcarrier $k$ is modeled as a complex matrix
$\mathbf{H}_{k}\in\mathbb{C}^{N_{r}\times2N_{t}}$, where $2N_{t}$
accounts for the dual-polarization configuration of the transmitting
antennas. For a wideband orthogonal frequency division multiplexing
(OFDM) system employing $N_{f}$ subcarriers, the overall channel
can be represented as a three-dimensional tensor:
\[
\mathcal{H}=\left[\mathbf{H}_{1},\mathbf{H}_{2},...,\mathbf{H}_{N_{f}}\right]\in\mathbb{C}^{N_{r}\times2N_{t}\times N_{f}}
\]
For FDD systems, the BS obtains CSI by sending downlink pilots and
receiving feedback from the UE. It is assumed that the UE has obtained
perfect CSI via pilot-based estimation and this work concentrates
on the design of the feedback scheme. 

In our proposed feedback scheme, compression and feedback are performed
independently for each receiving antenna slice. Specifically, for
the $i$-th receiving antenna, the UE applies a compression operator
$\mathcal{F}\left(\cdot\right)$ to its corresponding channel slice
$\mathcal{H}_{i,:,:}\in\mathbb{C}^{2N_{t}\times N_{f}}$, producing
a low-dimensional compressed representation: 
\[
\mathbf{z}_{i}=\mathcal{F}\left(\mathcal{H}_{i,:,:}\right)\in\mathbb{C}^{d}
\]
where $d\ll2N_{t}N_{f}$ denotes the number of feedback complex numbers.
The compressed CSI vector $\mathbf{z}_{i}$ is then transmitted to
the BS. At the BS, the received feedback is decoded using a reconstruction
function $\mathcal{G}\left(\cdot\right)$ to obtain an estimate of
the original channel slice:
\[
\hat{\mathcal{H}}_{i,:,:}=\mathcal{G}\left(\mathbf{z}_{i}\right)\in\mathbb{C}^{2N_{t}\times N_{f}}
\]
By processing all $N_{r}$ slices in parallel, the BS reconstructs
the full channel tensor as
\[
\hat{\mathcal{H}}=\left[\hat{\mathcal{H}}_{1,:,:},\hat{\mathcal{H}}_{2,:,:},...,\hat{\mathcal{H}}_{N_{r},:,:}\right]\in\mathbb{C}^{N_{r}\times2N_{t}\times N_{f}}
\]

With $\hat{\mathcal{H}}$, the BS can perform the Singular Value Decomposition
(SVD)-based precoding to assist downlink communication\textcolor{blue}{\ }\cite{SVD_precoding}.
Specifically, for the $R$-th layer (data stream), the precoder is
chosen as the eigenvector $\mathbf{v}_{R,k}$ associated with the
$R$-th largest eigenvalue of $\hat{\mathbf{H}}_{k}^{H}\hat{\mathbf{H}}_{k}$
, where $\hat{\mathbf{H}}_{k}$ denotes the reconstructed channel.
By collecting the eigenvectors across all $N_{f}$ subbands, the precoding
matrix$\mathbf{W}_{R}$ for layer $R$ can be formed as follows:
\[
\mathbf{W}_{R}=\left[\mathbf{v}_{R,1},...,\mathbf{v}_{R,N_{f}}\right]\in\mathbb{C}^{2N_{t}\times N_{f}}
\]

\section{LI-Based Compression in the Frequency Domain\label{sec:3}}

This section presents the LI-based compression in the frequency domain
which consists of two steps. First, an LI-based CSI interpolation
model inspired by the standard Loewner matrix approach to the generalized
realization problem is developed. Then, we introduce the order reduction
for the LI model, along with a structural simplification of the bases
to further reduce the feedback overhead and facilitate the subsequent
spatial compression. 

\subsection{LI-Based CSI Interpolation Model}

The LI method establishes a direct connection between data-driven
modeling and the state-space realization of linear time-invariant
(LTI) systems from its frequency-domain sampling data and has been
proved to be suitable for solving massive-port problems\ \cite{LI_1_lefteriu2009modeling}\cite{LI_2_MAYO2007634}\cite{LI_3_benner2021system}.
Specifically, given a set of frequency-domain samples
\begin{equation}
\left\{ \left(s_{i},\mathbf{H}(s_{i})\right):s_{i}\in\mathbb{R},\mathbf{H}(s_{i})\in\mathbb{C}^{m\times n}\right\} 
\end{equation}
 from an unknown system, the Loewner framework can construct a system
descriptor:
\begin{equation}
\mathbf{\Sigma}\triangleq\left(\mathbf{E},\mathbf{A},\mathbf{B},\mathbf{C}\right)
\end{equation}
whose transfer function is as:\vspace{-0.3cm}
\begin{equation}
\mathbf{H}\left(s\right)=\mathbf{C}\left(s\mathbf{E}-\mathbf{A}\right)^{-1}\mathbf{B}
\end{equation}

\vspace{-0.25cm}
Inspired by the success of the LI framework in solving system realization
problems for dynamic systems, we extend this concept to the frequency-domain
compression of the CSI matrix. This extension of the LI framework
from LTI system modeling to the frequency-domain representation and
compression of CSI matrices is conceptually sound. In both cases,
the core task is to reconstruct or represent a parametric system from
its frequency-sampled data, whether it be a classical state-space
model or a high-dimensional wireless channel characterized by its
CSI. By interpreting the CSI matrices as frequency-response data,
the proposed method leverages the Loewner-based modeling mechanism
to capture the underlying correlation in the frequency domain and
obtain a compact representation. This analogy allows the CSI compression
problem to be reformulated as a data-driven realization task, enabling
accurate reconstruction at the receiver while significantly reducing
feedback overhead.

\begin{figure}[tbh]
\begin{centering}
\includegraphics[width=6.5cm,totalheight=7cm]{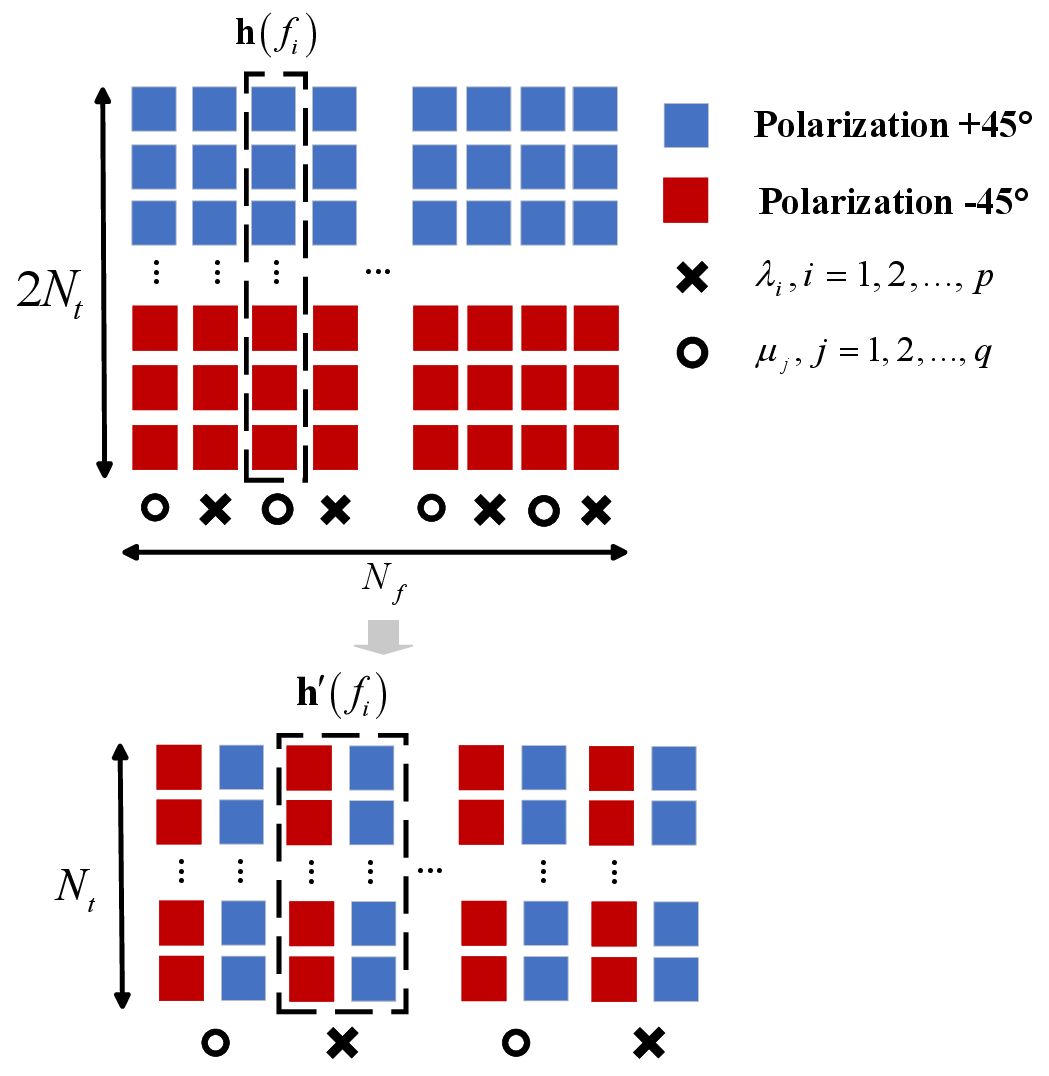}
\par\end{centering}
\caption{The LI-based CSI interpolation model\label{fig:1}.}
\end{figure}

We regard the original CSI slice matrix $\mathbf{H}=\mathcal{H}_{i,:,:}\in\mathbb{C}^{2N_{t}\times N_{f}}$
as a set of samples, as defined in \eqref{eq:4}. Each column of $\mathbf{H}=\mathcal{H}_{i,:,:}\in\mathbb{C}^{2N_{t}\times N_{f}}$
, indexed by $f_{i}$, represents a potential sample point with the
corresponding column vector $\mathbf{h}\left(f_{i}\right)\in\mathbb{C}^{2N_{t}\times1}$
serving as the associated sample value. The total number of sample
points is denoted by $N$ , which satisfies $1\leq N\leq N_{f}$.
\begin{equation}
\mathcal{S}=\left\{ f_{i},\mathbf{h}\left(f_{i}\right),i=1,2,...,N\right\} \label{eq:4}
\end{equation}

Each sample vector $\mathbf{h}\left(f_{i}\right)\in\mathbb{C}^{2N_{t}\times1}$
is first reshaped to $\mathbf{h}^{\prime}\left(f_{i}\right)\in\mathbb{C}^{N_{t}\times2}$
according to the dual-polarization configuration of the transmitting
antennas. The set $S$ is then divided into left and right interpolation
subsets respectively, denoted as $\mathcal{S}=\varLambda\cup M$,
leading to\ \eqref{eq:5}. The CSI interpolation model is constructed
as illustrated in Fig.\ \ref{fig:1}.
\begin{equation}
\begin{array}{c}
\varLambda=\left\{ \lambda_{i},\mathbf{h}^{\prime}\left(\lambda_{i}\right),i=1,2,...,p\right\} \\
M=\left\{ \mu_{j},\mathbf{h}^{\prime}\left(\mu_{j}\right),j=1,2,...,q\right\} \\
p+q=N
\end{array}\label{eq:5}
\end{equation}

Using the left and right sampling sets, we construct the matrices
$\mathbf{V}$ and $\mathbf{W}$ as follows \eqref{eq:6}:
\begin{equation}
\begin{aligned}\mathbf{V}= & \left[\begin{array}{c}
\mathbf{h}^{\prime}\left(\lambda_{1}\right)\\
\vdots\\
\mathbf{h}^{\prime}\left(\lambda_{p}\right)
\end{array}\right]\in\mathbb{C}^{pN_{t}\times2}\\
\mathbf{W}= & \left[\begin{array}{ccc}
\mathbf{h}^{\prime}\left(\mu_{1}\right) & \cdots & \mathbf{h}^{\prime}\left(\mu_{q}\right)\end{array}\right]\in\mathbb{C}^{N_{t}\times2q}
\end{aligned}
\label{eq:6}
\end{equation}

Subsequently, the Loewner matrix $\text{\ensuremath{\mathbb{L}}}\in\mathbb{C}^{pN_{t}\times2q}$
and the shifted Loewner matrix $\mathbb{L_{\sigma}}\in\mathbb{C}^{pN_{t}\times2q}$
are formed as in \eqref{eq:7} and \eqref{eq:8}. Both $\mathbb{L}$
and $\mathbb{L}_{\sigma}$ are block matrices of size $p\times q$,
with each block being a $N_{t}\times2$ submatrix: 
\begin{equation}
\mathbb{L}=\left[\begin{array}{c}
\frac{\mathbf{h}^{\prime}\left(\lambda_{i}\right)-\mathbf{h}^{\prime}\left(\mu_{j}\right)}{\lambda_{i}-\mu_{j}}\end{array}\right]_{\begin{array}{c}
\forall i=1,...,p\\
\forall j=1,...,q
\end{array}}\label{eq:7}
\end{equation}
\begin{equation}
\mathbb{L_{\sigma}}=\left[\begin{array}{c}
\frac{\lambda_{i}\mathbf{h}^{\prime}\left(\lambda_{i}\right)-\mu_{j}\mathbf{h}^{\prime}\left(\mu_{j}\right)}{\lambda_{i}-\mu_{j}}\end{array}\right]_{\begin{array}{c}
\forall i=1,...,p\\
\forall j=1,...,q
\end{array}}\label{eq:8}
\end{equation}

According to the above definition, we have $\left(\mathbb{L}_{\sigma}-\lambda_{i}\mathbb{L}\right)_{i,j}=\mathbf{h}^{\prime}\left(\mu_{j}\right)$.
Let $\mathbb{I}_{i}=\left[\mathbf{0}\cdots\mathbf{I}_{i}\in\mathbb{R}^{N_{t}\times N_{t}}\cdots\mathbf{0}\right]\in\mathbb{R}^{N_{t}\times pN_{t}}$
denote a $1\times p$ block matrix where $\mathbf{I}_{i}$ is an identity
matrix at block $i$, we have $\mathbb{I}_{i}\mathbf{V}=\mathbf{h}^{\prime}\left(\lambda_{i}\right)$
and $\mathbb{I}_{i}\left(\mathbb{L}_{\sigma}-\lambda_{i}\mathbb{L}\right)=\mathbf{W}$
for $\forall i=1,...,p$. If $\left(\mathbb{L}_{\sigma}-\lambda_{i}\mathbb{L}\right)$
is invertible, we have $\mathbf{h}^{\prime}\left(\lambda_{i}\right)=\mathbb{I}_{i}\mathbf{V}=\mathbf{W}\left(\mathbb{L}_{\sigma}-\lambda_{i}\mathbb{L}\right)^{-1}\mathbf{V}$.
Consequently, the interpolation function is defined as \eqref{eq:11}:
\begin{equation}
\mathbf{g}\left(f\right)=\mathbf{W}\left(\mathbb{L}_{\sigma}-f\mathbb{L}\right)^{-1}\mathbf{V}\in\mathbb{C}^{N_{t}\times2}\label{eq:11}
\end{equation}

The above proves that $\mathbf{g}\left(f\right)$ guarantees the perfect
recovery at the sample point $\lambda_{i}$, i.e. $\mathbf{g}\left(\lambda_{i}\right)=\mathbf{h}^{\prime}\left(\lambda_{i}\right),i=1,2,...,p$.
It is also easy to prove the perfect recovery for the other sample
set: $\mathbf{g}\left(\mu_{j}\right)=\mathbf{h}^{\prime}\left(\mu_{j}\right),j=1,2,...,q$.\textcolor{red}{{} }

\subsection{MOR for CSI Interpolation}

To prevent an excessive feedback overhead, the MOR in the LI framework~\cite{MOR_approximate_2}
is applied to $\left\{ \mathbf{W},\mathbf{V},\mathbb{L},\mathbb{L}_{\sigma}\right\} $.\textcolor{red}{{}
}We first select a specific sample point $\lambda_{i}$ and perform
a SVD decomposition of $\mathbb{L}_{\sigma}-\lambda_{i}\mathbb{L}$
to obtain the corresponding projection matrices as \eqref{eq:12}.
In practice, $\lambda_{i}$ is chosen as the midpoint of the sampling
set.
\begin{equation}
\begin{array}{c}
\mathbb{L}_{\sigma}-\lambda_{i}\mathbb{L}=\mathbf{Y}\mathbf{\Sigma}_{i}\mathbf{X}^{H}\\
\mathbf{Y}\in\mathbb{C}^{pN_{t}\times pN_{t}},\mathbf{X}\in\mathbb{C}^{2q\times2q},\mathbf{\Sigma}_{i}\in\mathbb{R}^{pN_{t}\times2q}
\end{array}\label{eq:12}
\end{equation}

After applying the projection matrices $\mathbf{X},\mathbf{Y}$ to
$\left\{ \mathbf{V},\mathbf{W},\mathbb{L},\mathbb{L}_{\sigma}\right\} $,
the interpolation function can be written as \eqref{eq:13}:
\begin{equation}
\begin{array}{c}
\mathbf{g}\left(f\right)=\left(\mathbf{W}\mathbf{X}\right)\left(\mathbf{X}^{H}\left(\mathbb{L}_{\sigma}-f\mathbb{L}\right)^{-1}\mathbf{Y}\right)\left(\mathbf{Y}^{H}\mathbf{V}\right)\\
=\left(\mathbf{W}\mathbf{X}\right)\left(\mathbf{Y}^{H}\mathbb{L}_{\sigma}\mathbf{X}-f\mathbf{Y}^{H}\mathbb{L}\mathbf{X}\right)^{-1}\left(\mathbf{Y}^{H}\mathbf{V}\right)
\end{array}\label{eq:13}
\end{equation}

If the projection matrices $\mathbf{X}$ and $\mathbf{Y}$ are suitably
truncated, the feedback overhead can be reduced with only a slight
decrease in the reconstruction accuracy. Specifically, we preserve
the singular vectors corresponding to the largest $r_{f}$ singular
values of $\mathbf{X},\mathbf{Y}$ to construct the truncated matrices
$\mathbf{X}_{p},\mathbf{Y}_{p}$ where $r_{f}\ll\left\{ pN_{t},2q\right\} $.
This leads to the reduced-order interpolation function $\mathbf{g}\left(f\right)\approx\mathbf{C}_{1}\left(f\mathbf{E}_{1}-\mathbf{A}_{1}\right)^{-1}\mathbf{B}_{1}\in\mathbb{C}^{N_{t}\times2}$
where $\mathbf{A}_{1},\mathbf{B}_{1},\mathbf{C}_{1},\mathbf{E}_{1}$
are defined in \eqref{eq:14}. The parameter $r_{f}$ represents the
frequency order and is critical in the CSI interpolation model, striking
a balance between recovery accuracy and feedback overhead.
\begin{equation}
\begin{cases}
\mathbf{C}_{1}\triangleq & \mathbf{WX}_{p}\in\mathbb{C}^{N_{t}\times r_{f}}\\
\mathbf{E}_{1}\triangleq & -\mathbf{Y}_{p}^{H}\mathbb{L}\mathbf{X}_{p}\mathbf{\in\mathbb{C}}^{r_{f}\times r_{f}}\\
\mathbf{A}_{1}\triangleq & -\mathbf{Y}_{p}^{H}\mathbb{L_{\sigma}}\mathbf{X}_{p}\mathbf{\in\mathbb{C}}^{r_{f}\times r_{f}}\\
\mathbf{B}_{1}\triangleq & \mathbf{Y}_{p}^{H}\mathbf{V}\in\mathbb{C}^{r_{f}\times2}
\end{cases}\label{eq:14}
\end{equation}

Since the dimension of $\mathbf{A}_{1},\mathbf{B}_{1},\mathbf{C}_{1},\mathbf{E}_{1}$
is still large, the feedback overhead can be further reduced by turning
$\mathbf{E}_{1}$ to identity matrix with SVD $\mathbf{E}_{1}=\mathbf{U}_{e}\mathbf{\Sigma}_{e}\mathbf{V}_{e}^{H}\in\mathbb{C}^{r_{f}\times r_{f}}\ast\mathbb{R}^{r_{f}\times r_{f}}\ast\mathbb{C}^{r_{f}\times r_{f}}$
and obtain \eqref{eq:15}. 
\begin{equation}
\begin{cases}
\mathbf{C}_{2}\triangleq & \mathbf{C}_{1}\mathbf{V}_{e}\sqrt{\mathbf{\Sigma}_{e}^{-1}}\in\mathbb{C}^{N_{t}\times r_{f}}\\
\mathbf{A}_{2}\triangleq & \sqrt{\mathbf{\mathbf{\Sigma}}_{e}^{-1}}\mathbf{U}_{e}^{H}\mathbf{A}_{1}\mathbf{V}_{e}\sqrt{\mathbf{\mathbf{\Sigma}}_{e}^{-1}}\in\mathbb{C}^{r_{f}\times r_{f}}\\
\mathbf{B}_{2}\triangleq & \sqrt{\mathbf{\mathbf{\mathbf{\Sigma}}}_{e}^{-1}}\mathbf{U}_{e}^{H}\mathbf{B}_{1}\in\mathbb{C}^{r_{f}\times2}
\end{cases}\label{eq:15}
\end{equation}

Then by diagonalizing $\mathbf{A}_{2}$ with SVD $\mathbf{A}_{2}=\mathbf{U}_{a}\mathbf{\Sigma}_{a}\mathbf{U}_{a}^{-1}\in\mathbb{C}^{r_{f}\times r_{f}}\ast\mathbb{\mathbb{C}}^{r_{f}\times r_{f}}\ast\mathbb{C}^{r_{f}\times r_{f}}$,
the interpolation function can be further written as $\mathbf{g}\left(f\right)\approx\mathbf{C}_{3}\left(f\mathbf{I}-\mathbf{A}_{3}\right)^{-1}\mathbf{B}_{3}$
where the new matrices $\left\{ \mathbf{A}_{3},\mathbf{B}_{3},\mathbf{C}_{3}\right\} $,
as expressed in \eqref{eq:16}, implicitly absorb the normalization
and diagonalization transformations introduced in the previous steps.
\begin{equation}
\begin{cases}
\mathbf{C}_{3}\triangleq & \mathbf{C}_{2}\mathbf{U}_{a}\in\mathbb{C}^{N_{t}\times r_{f}}\\
\mathbf{A}_{3}\triangleq & \mathbf{\Sigma}_{a}\in\mathbb{C}^{r_{f}\times r_{f}}\\
\mathbf{B}_{3}\triangleq & \mathbf{U}_{a}^{-1}\mathbf{B}_{2}\in\mathbb{C}^{r_{f}\times2}
\end{cases}\label{eq:16}
\end{equation}

The dimension of matrices $\left\{ \mathbf{A}_{3},\mathbf{B}_{3},\mathbf{C}_{3}\right\} $
is closely related to the order $r_{f}$ where $r_{f}\ll\left\{ pN_{T},2q\right\} $.
Given any frequency index $f_{p}\in\left\{ 1,2,...,N_{f}\right\} $,
the BS can recover $\widehat{\mathbf{H}}\left(f_{p}\right)\in\mathbb{C}^{N_{t}\times2}$
with \eqref{eq:17}, which is the channel coefficients of the two
polarization at subcarrier $f_{p}$. 
\begin{equation}
\widehat{\mathbf{H}}\left(f_{p}\right)=\mathbf{C}_{3}\left(f_{p}\mathbf{I}-\mathbf{A}_{3}\right)^{-1}\mathbf{B}_{3}\label{eq:17}
\end{equation}

\begin{figure*}[t]
\begin{centering}
\includegraphics[width=16cm,totalheight=3.8cm]{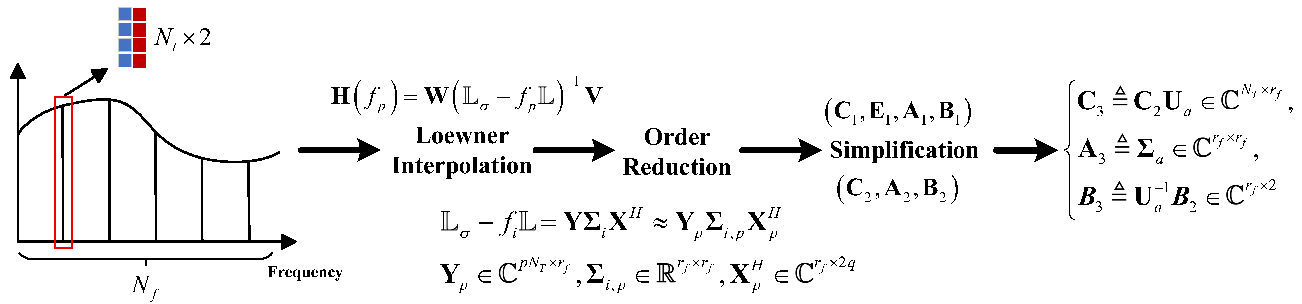}
\par\end{centering}
\caption{The illustration of the LI-based CSI compression scheme.\label{fig:2}}

\end{figure*}

The proposed LI-based CSI compression scheme for the frequency domain
is summarized in Fig.\ \ref{fig:2}. By integrating the LI with MOR
and further simplifying the basis, the high-dimensional CSI data are
efficiently projected onto a compact and well-conditioned subspace
and can be effectively represented by a set of descriptor $\left\{ \mathbf{A}_{3},\mathbf{B}_{3},\mathbf{C}_{3}\right\} $.
After the compression in the frequency domain, the feedback overhead
(complex numbers) is $\left(3r_{f}+r_{f}\times N_{t}\right)$.

\section{Rateless Auto-encoder-Based Compression in the Spatial Domain\label{sec:4}}

In this section, we introduce the rateless auto-encoder-based compression
scheme for the spatial domain. We begin by introducing the general
DL-based compression framework. Then, we develop a cascaded framework
for the spatial domain compression architecture for $\mathbf{C}_{3}$
(one of the LI bases) which incorporates both the prevention of error
amplification and sparsity enhancement. Finally, we introduce the
implementation of the rateless auto-encoder which is called LI-MORNet
that supports the variable-length feedback. 

\subsection{DL-Based Compression}

We consider DL-based compression in the spatial domain by feeding
the target matrix $\mathbf{C}$ into an auto-encoder for further compression.
Specifically, we aim to design an encoder: 
\begin{equation}
\mathbf{v}=f_{en}(\mathbf{C},\theta_{en})
\end{equation}
which maps $\mathbf{C}\in\mathbb{C}^{m\times n}$ into an $M$-dimensional
real-vector codeword $\mathbf{v}\in\mathbb{R}^{M}$, where $\theta_{en}$
denotes the parameters of the encoder. The CR in the spatial domain
is therefore defined as:
\begin{equation}
CR_{s}=\frac{M}{2mn}\label{eq:19}
\end{equation}

In addition, we also design a decoder that reconstructs $\hat{\mathbf{C}}$
from the received codeword $\mathbf{v}$. Considering that the feedback
is based on a quantized representation $\mathbf{v}_{q}=\mathcal{Q}\left(\mathbf{v}\right)$,
$\hat{\mathbf{C}}$ recovered at the BS is given by:
\begin{equation}
\hat{\mathbf{C}}=f_{de}(\mathcal{D}\left(\mathbf{v}_{q}\right),\varphi_{de})
\end{equation}
where $\mathcal{Q}\left(\cdot\right)$ and $\mathcal{D}\left(\cdot\right)$
denote the quantization and dequantization functions respectively.
$\varphi_{de}$ represents the parameters of the decoder, and $\mathbf{v}_{q}$
is the quantized codeword. 

\subsection{Spatial Compression in the Cascaded Framework\label{subsec:4-2}}

After the compression in the frequency domain, we obtain $\left\{ \mathbf{A}_{3}\in\mathbb{C}^{r_{f}\times r_{f}},\mathbf{B}_{3}\in\mathbb{C}^{r_{f}\times2},\mathbf{C}_{3}\in\mathbb{C}^{N_{t}\times r_{f}}\right\} $
where $r_{f}$ denotes the preserved order and is typically small.
Consequently, $\mathbf{A}_{3}$ and $\mathbf{B}_{3}$ are low-dimensional
matrices. In contrast, $\mathbf{C}_{3}\in\mathbb{C}^{N_{t}\times r_{f}}$
contains redundant information in the spatial domain with dimension
$N_{t}$ and thus can be further compressed. 

However, directly feeding $\mathbf{C}_{3}$ into the auto-encoder
leads to a critical issue: the reconstruction error $\bigtriangleup\mathbf{C}_{3}=\hat{\mathbf{C}_{3}}-\mathbf{C}_{3}=f_{de}(f_{en}(\mathbf{C}_{3},\theta_{en}),\varphi_{de})-\mathbf{C}_{3}$
introduced by the auto-encoder will be subsequently amplified during
the subsequent recovery in the frequency domain via the interpolation
function. This error propagation arises from the cascaded structure
of the two-step compression framework. In other words, $\bigtriangleup\mathbf{C}_{3}$
is not only preserved but further amplified when passed through the
LI-based reconstruction module in the frequency domain. The following
provides a detailed explanation of this error-amplification mechanism.

Consider the recovery interpolation function as \eqref{eq:17} with
the reconstructed $\hat{\mathbf{C}_{3}}$ while assuming $\left\{ \mathbf{A}_{3},\mathbf{B}_{3}\right\} $
are perfectly recovered. The reconstructed CSI $\hat{\mathbf{H}}_{s}$
at the sampled frequency index $\left\{ f_{1},f_{2},...,f_{N}\right\} $
is expressed as \eqref{eq:21}. In the following discussions, $\mathbf{H}_{s}$
refers exclusively to the channel matrix associated with the sampled-frequency
indices, while those without the subscript $s$ correspond to the
full-subband counterparts.
\begin{equation}
\begin{alignedat}{1}\hat{\mathbf{H}}_{s}= & \left[\mathbf{h}^{\prime}\left(f_{1}\right),\mathbf{h}^{\prime}\left(f_{2}\right),...,\mathbf{h}^{\prime}\left(f_{N}\right)\right]\\
= & \hat{\mathbf{C}}_{3}\left[\left(f_{1}\mathbf{I}-\mathbf{A}_{3}\right)^{-1}\mathbf{B}_{3}...\left(f_{N}\mathbf{I}-\mathbf{A}_{3}\right)^{-1}\mathbf{B}_{3}\right]
\end{alignedat}
\label{eq:21}
\end{equation}

For notational compactness, define $\mathbf{Y}$ as \eqref{eq:22}
where $r_{f}<2N$. 
\begin{equation}
\mathbf{Y}=\left[\left(f_{1}\mathbf{I}-\mathbf{A}_{3}\right)^{-1}\mathbf{B}_{3}...\left(f_{N}\mathbf{I}-\mathbf{A}_{3}\right)^{-1}\mathbf{B}_{3}\right]\in\mathbb{C}^{r_{f}\times2N}\label{eq:22}
\end{equation}

Then, the squared $\mathcal{F}$-norm of the reconstructed error $\triangle\mathbf{H}_{s}$
can be written as: 
\begin{equation}
\begin{alignedat}{1}\left\Vert \triangle\mathbf{H}_{s}\right\Vert _{F}^{2} & =\left\Vert \mathbf{H}_{s}-\left(\mathbf{C}_{3}+\bigtriangleup\mathbf{C}_{3}\right)\mathbf{Y}\right\Vert _{F}^{2}\\
=\left\Vert \bigtriangleup\mathbf{C}_{3}\mathbf{Y}\right\Vert _{F}^{2} & =tr\left(\left(\bigtriangleup\mathbf{C}_{3}\mathbf{Y}\right)^{H}\left(\bigtriangleup\mathbf{C}_{3}\mathbf{Y}\right)\right)\\
 & =tr\left(\mathbf{Y}^{H}\bigtriangleup\mathbf{C}_{3}^{H}\bigtriangleup\mathbf{C}_{3}\mathbf{Y}\right)\\
 & =tr\left(\bigtriangleup\mathbf{C}_{3}^{H}\bigtriangleup\mathbf{C}_{3}\mathbf{Y}\mathbf{Y}^{H}\right)
\end{alignedat}
\label{eq:23}
\end{equation}

As shown in \eqref{eq:23}, the reconstruction error at the sampled
points depends not only on $\bigtriangleup\mathbf{C}_{3}$ which is
introduced by the compression in the spatial domain but is further
amplified by $\mathbf{Y}\mathbf{Y}^{H}$. To eliminate this amplification
effect, we aim to transform $\mathbf{Y}$ into an orthogonal matrix
such that $\mathbf{Y}\mathbf{Y}^{H}=\mathbf{I}$, thereby eliminating
the error-scaling effect.

To achieve this, we perform the thin SVD to $\mathbf{Y}$ as shown
in \eqref{eq:24}.
\begin{equation}
\begin{array}{c}
\mathbf{Y}=\mathbf{U}_{Y}\mathbf{\Sigma}_{Y}\mathbf{V}_{Y}^{H}\\
\mathbf{U}_{Y}\in\mathbb{C}^{r_{f}\times r_{f}},\mathbf{\Sigma}_{Y}\in\mathbb{R}^{r_{f}\times r_{f}},\mathbf{V}_{Y}^{H}\in\mathbb{C}^{r_{f}\times2N}
\end{array}\label{eq:24}
\end{equation}

We then replace $\mathbf{Y}$ with $\mathbf{V}_{Y}^{H}$ and absorb
the remaining factors $\mathbf{U}_{Y}$ and $\mathbf{\mathbf{\Sigma}}_{Y}$
into $\mathbf{C}_{3}$, yielding a transformed basis as: 
\begin{equation}
\mathbf{C}_{4}\triangleq\mathbf{C}_{3}\mathbf{U}_{Y}\mathbf{\Sigma}_{Y}\in\mathbb{C}^{N_{t}\times r_{f}}\label{eq:25}
\end{equation}

The reconstruction error of this adjusted scheme is given in \eqref{eq:26}.
Compared to \eqref{eq:23}, it is obvious that the error-amplification
effect of $\mathbf{Y}\mathbf{Y}^{H}$ has been fully removed.
\begin{equation}
\begin{alignedat}{1}\left\Vert \triangle\mathbf{H}_{s}\right\Vert _{F}^{2}= & \left\Vert \mathbf{H}_{s}-\left(\mathbf{C}_{4}+\bigtriangleup\mathbf{C}_{4}\right)\mathbf{V}_{Y}^{H}\right\Vert _{F}^{2}\\
=\left\Vert \Delta\mathbf{C}_{4}\mathbf{V}_{Y}^{H}\right\Vert _{F}^{2} & =\left\Vert \Delta\mathbf{C}_{4}\right\Vert _{F}^{2}
\end{alignedat}
\label{eq:26}
\end{equation}

To facilitate the auto-encoder-based compression in the spatial domain,
we further enhance the sparsity of $\mathbf{C}_{4}$ with DFT. Accordingly,
the matrix fed into the auto-encoder becomes \eqref{eq:27}:
\begin{equation}
\mathbf{C}_{5}=\mathbf{F}_{a}\mathbf{C}_{4}\in\mathbb{C}^{N_{t}\times r_{f}}\label{eq:27}
\end{equation}
where $\mathbf{F}_{a}\in\mathbb{C}^{N_{t}\times N_{t}}$ denotes the
unitary DFT matrix with $\left(\mathbf{F}_{a}\right)_{i,j}=\frac{1}{\sqrt{N_{t}}}e^{-j2\pi\left(i-1\right)\left(j-1\right)/N_{t}}$.
Since no truncation is performed on the sparse matrix $\mathbf{C}_{5}$,
the sparsity enhancement step does not introduce additional reconstruction
errors as proved in \eqref{eq:28}. 
\begin{equation}
\begin{alignedat}{1}\left\Vert \Delta\mathbf{C}_{4}\right\Vert _{F}^{2}= & \left\Vert \hat{\mathbf{C}}_{4}-\mathbf{C}_{4}\right\Vert _{F}^{2}=\left\Vert \mathbf{F}_{a}^{H}\hat{\mathbf{C}}_{5}-\mathbf{F}_{a}^{H}\mathbf{C}_{5}\right\Vert _{F}^{2}\\
= & \left\Vert \mathbf{F}_{a}^{H}\Delta\mathbf{C}_{5}\right\Vert _{F}^{2}=\left\Vert \Delta\mathbf{C}_{5}\right\Vert _{F}^{2}
\end{alignedat}
\label{eq:28}
\end{equation}

Assuming that the BS has obtained the recovered $\hat{\mathbf{C}}_{5}$
from the decoder, the interpolation of $\widehat{\mathbf{H}}\left(f_{p}\right)$
(defined in \eqref{eq:17}) at any given frequency index $f_{p}$
requires to reconstruct $\hat{\mathbf{C}}_{3}$ from $\hat{\mathbf{C}}_{5}$.
This reconstruction corresponds the reverse process of \eqref{eq:27}
and \eqref{eq:25} which is given by \eqref{eq:29}: 
\begin{equation}
\hat{\mathbf{C}}_{3}=\mathbf{F}_{a}^{H}\hat{\mathbf{C}}_{5}\hat{\mathbf{\Sigma}}_{Y}^{-1}\mathbf{\hat{U}}_{Y}^{H}\in\mathbb{C}^{N_{t}\times r_{f}}\label{eq:29}
\end{equation}

$\mathbf{\hat{U}}_{Y}$ and $\mathbf{\hat{\Sigma}}_{Y}$ are obtained
from the SVD of $\hat{\mathbf{Y}}$ with the feedback $\left\{ \mathbf{\hat{A}}_{3},\mathbf{\hat{B}}_{3}\right\} $
and the sample index $\left\{ f_{1},f_{2},...,f_{N}\right\} $ as
defined in \eqref{eq:22}. In practice, uniform sampling is adopted
for the LI in the frequency domain. Consequently, the BS and the UE
only need to maintain a consistent sampling interval, and there is
no need to explicitly feed back the indices of the sampled frequency
points. 

\subsection{Implementation of the Rateless Auto-Encoder}

\begin{figure*}[t]
\begin{centering}
\includegraphics[width=19cm,totalheight=6cm]{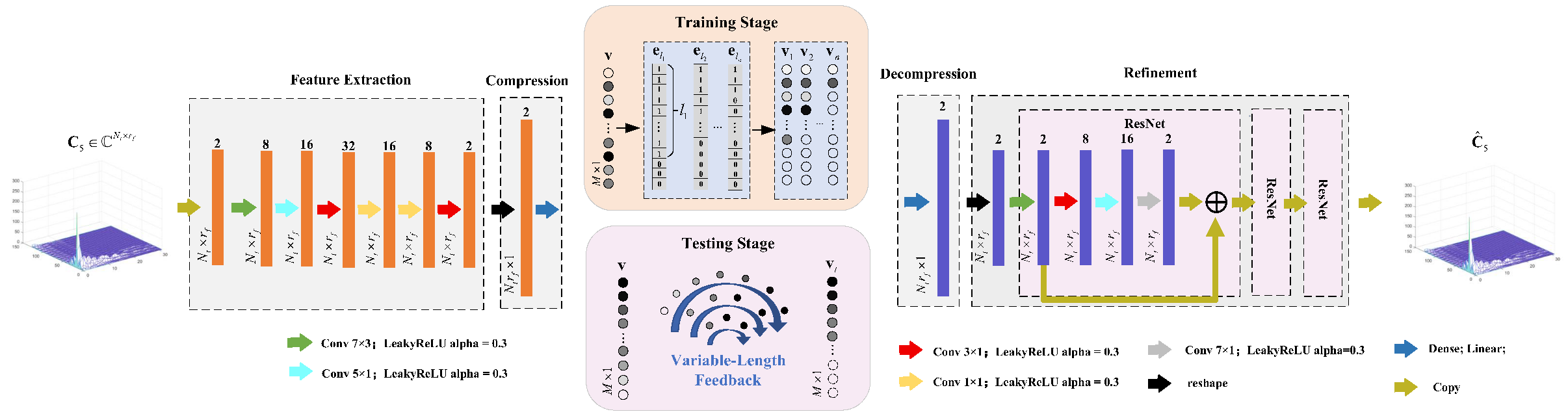}
\par\end{centering}
\caption{The illustration of the proposed LI-MORnet.\label{fig:3} }
\end{figure*}

The LI-MORNet that achieves the flexible compression with variable-length
codewords is implemented through two steps as shown in Fig.\ \ref{fig:3}.
First, we introduce the fixed CR module based on a CNN-based auto-encoder
architecture. In the feature extraction module of the encoder, we
adopt a cascaded convolutional architecture that starts with several
large-sized kernels and gradually transitions to smaller kernels.
The number of channels gradually increases with the number of convolutional
layers and then decreases. The strides of all feature extraction layers
are $\left(1,1\right)$, ensuring that the dimension of $\mathbf{C}_{5}$
remains unchanged while the network focuses on learning its feature
representation. The activation functions are leaky rectified linear
unit (Leaky ReLU) for all the convolutional layers. A fully connected
(FC) layer is employed after the feature extraction to generate an
initial compressed codeword whose length is $M\times1$ (fixed CR).
In the decoder, following the decompression layer, three residual
blocks are used to further refine the reconstructed $\hat{\mathbf{C}}_{5}$. 

Then, we introduce the flexible compression mechanism, which is applied
after the feature compression module using a set of random binary
masks. During the training phase, a set of masks as: 
\[
\left\{ \mathbf{e}_{l_{1}},\mathbf{e}_{l_{2}},...,\mathbf{e}_{l_{n}}\right\} ,\mathbf{e}_{l_{i}}\triangleq\left[\underset{l_{i}}{\underbrace{1,1,...,1}},0,0,...,0\right]^{T}\in\mathbb{R}^{M\times1}
\]
is applied to the initial codeword $\mathbf{v}$ to generate multiple
virtually truncated vectors:
\begin{equation}
\left\{ \mathbf{v}_{1},\mathbf{v}_{2},...,\mathbf{v}_{n}\right\} ,\mathbf{v}_{i}=\mathbf{v}\odot\mathbf{e}_{l_{i}}\in\mathbb{R}^{M\times1}
\end{equation}
To ensure diversity in the truncation degrees, the length $l_{i}$
of each mask is randomly sampled within a predefined interval as $l_{i}\sim\mathcal{U}\left(L_{i}^{min},L_{i}^{max}\right).$
Furthermore, the $n$ intervals $\left\{ \left[L_{1}^{min},L_{1}^{max}\right],\left[L_{2}^{min},L_{2}^{max}\right],...,\left[L_{n}^{min},L_{n}^{max}\right]\right\} $
are designed to be mutually non-overlapping and equal in length, i.e.
$L_{i}^{max}-L_{i}^{min}=\Delta L,\forall i\in\left\{ 1,2,...,n\right\} $.
Each masked codewords $\mathbf{v}_{i}$ is then fed into the decoder
to obtain a set of reconstructed matrices:
\begin{equation}
\left\{ \mathbf{\hat{C}}_{5,1},\mathbf{\hat{C}}_{5,2},...,\mathbf{\hat{C}}_{5,n}\right\} ,\mathbf{\hat{C}}_{5,i}=f_{de}\left(\mathbf{v}_{i},\varphi_{de}\right)
\end{equation}

For each reconstruction, we compute the corresponding mean square
error (MSE) which is defines as:
\begin{equation}
\mathcal{L}_{i}\left(\theta_{en},\varphi_{de};l_{i}\right)=\left\Vert \mathbf{C}_{5}-\mathbf{\hat{C}}_{5,i}\right\Vert _{F}^{2}\label{eq:33}
\end{equation}

We then take the weighted average of these MSE as the loss function
as :
\begin{equation}
\mathcal{L}\left(\theta_{en},\varphi_{de}\right)=\sum_{i=1}^{n}w_{i}\mathcal{L}_{i}\left(\theta_{en},\varphi_{de};l_{i}\right)\label{eq:34}
\end{equation}
Since the magnitudes of individual MSE losses $\mathcal{L}_{i}\left(\theta_{en},\varphi_{de};l_{i}\right)$
may vary significantly across different tasks or output levels $l_{i}$,
directly summing them could cause certain terms to dominate the overall
optimization process. Consequently, the weights $w_{i}$ are chosen
to adjust the individual loss terms so that they contribute at comparable
levels. The proposed LI-MORNet is trained using the adaptive moment
estimation (Adam) optimizer and a cosine annealing schedule for the
learning rate. With this training strategy, the compressed codeword
are automatically ordered by importance: the elements appearing earlier
in the codeword carry more critical feature information and thus receive
higher priority during feedback transmission. As a result, the receiver
does not need to obtain the entire codeword before initiating the
decoding process. During the test stage, after the encoder outputs
the codeword, the UE sequentially feeds back codewords of arbitrary
length. Upon receiving the codewords, the BS pads them to length $M$
before sending them to the decoder for reconstruction. Consequently,
the BS can progressively reconstruct $\hat{\mathbf{C}}_{5}$ using
only the first portion of the received codeword. This property makes
the proposed framework particularly suitable for scenarios involving
reception interruption or delayed decoding, which can be modeled as
a channel with tail erasures. The detail procedure of the LI-MORNet
is given in Algorithm\ \ref{alg:1}.

\begin{algorithm}[tbh]
\caption{The proposed LI-MORNet\label{alg:1}}

\textbf{Input:} Training set: $\mathcal{D}_{train}$; Testing set:
$\mathcal{D}_{test}$; Initial codeword length: $M$; Random mask
interval: $\left\{ \left[L_{1}^{min},L_{1}^{max}\right],\left[L_{2}^{min},L_{2}^{max}\right],...,\left[L_{n}^{min},L_{n}^{max}\right]\right\} $;
Maximum epoches: $E$; Test feedback length $l_{t}$, $\left(L_{1}^{min}\leq l_{t}\leq L_{n}^{max}\right)$;

\begin{algorithmic}[1]

\STATE\textbf{{[}Off-line training stage{]}:}

\STATE\textbf{Initialization:} $\theta_{en}^{0}$, $\varphi_{de}^{0}$\textcolor{blue}{.}

\FOR{$i=1\ldots E$}

\FOR{$\mathbf{C_{5}}$ in $\mathcal{D}_{train}$}

\FOR{$m=1\ldots n$}

\STATE Obtain codeword $\mathbf{v}$ from the encoder;

\STATE Randomly sampled mask length from the $m$-th interval $l_{m}\sim\mathcal{U}\left(L_{m}^{min},L_{m}^{max}\right)$

\STATE Apply mask $\mathbf{e}_{l_{m}}$ to $\mathbf{v}$ and obtain
virtually truncated vector $\mathbf{v}_{m}$;

\STATE Feed $\mathbf{v}_{m}$ to the decoder and obtain reconstructed
$\hat{\mathbf{C}}_{5,m}$;

\STATE Compute the MSE according to \eqref{eq:33}; 

\ENDFOR

\STATE Update $\theta_{en}^{i},\varphi_{de}^{i}$ via end-to-end
training with the loss defined in \eqref{eq:34};

\ENDFOR

\ENDFOR

\STATE\textbf{Output:} $\theta_{en}^{E},\varphi_{de}^{E}$ .

\STATE\textbf{{[}Online testing stage{]}:}

\FOR{$\mathbf{C_{5}}$ in $\mathcal{D}_{test}$}

\STATE Obtain codeword $\mathbf{v}$ from the encoder;

\STATE Apply mask $\mathbf{e}_{l_{t}}$ to the codeword $\mathbf{v}$
and obtain $\mathbf{v}_{t}$;

\STATE Feed $\mathbf{v}_{t}$ to the decoder and obtain $\mathbf{\hat{C}}_{5,t}$;

\ENDFOR

\STATE\textbf{Output:} $\mathbf{\hat{C}}_{5,t}$. 

\end{algorithmic}
\end{algorithm}

\section{Variable-Length Wideband CSI Feedback with Robust Quantization\label{sec:5}}

In this section, we develop a robust quantization scheme for the proposed
CSI compression framework. We first summarize the overall codewords
for feedback and design dedicated quantization schemes respectively
based on their impact on the reconstruction. We then outline the complete
procedure that integrates the quantization scheme into the end-to-end
reconstruction pipeline.

\subsection{Robust Quantization Algorithm\label{subsec:5-1}}

By combining the LI-based compression in the frequency domain (Section\ \ref{sec:3})
with the rateless auto-encoder-based compression in the spatial domain
(Section\ \ref{sec:4}), the resulting feedback codeword set consists
of:
\begin{equation}
\mathbf{z}=\left\{ \mathbf{A}_{3}\in\mathbb{C}^{r_{f}\times r_{f}},\mathbf{B}_{3}\in\mathbb{C}^{r_{f}\times2},\mathbf{v}_{l_{m}}\in\mathbb{R}^{l_{m}}\right\} .
\end{equation}

For the quantization of the complex-valued matrices $\mathbf{A}_{3}$
and $\mathbf{B}_{3}$, according to the transformation strategy designed
to suppress the error amplification in Section\ \ref{subsec:4-2},
$\mathbf{V}_{Y}$ must retain unitary. As defined in \eqref{eq:22},
$\mathbf{Y}$ is constructed from $\mathbf{A}_{3},\mathbf{B}_{3}$
and the set of sampled frequency indices $\left\{ f_{1},f_{2},...,f_{N}\right\} $.
Therefore, $\mathbf{A}_{3},\mathbf{B}_{3}$ need to be quantized with
sufficiently high precision to ensure that the reconstruction $\hat{\mathbf{V}}_{Y}$
satisfies $\hat{\mathbf{V}}_{Y}^{H}\hat{\mathbf{V}}_{Y}\approx\mathbf{I}_{r_{f}}$.
As a result, we perform uniform quantization to both the magnitude
and phase components of $\mathbf{A}_{3}$ and $\mathbf{B}_{3}$ with
high quantization bits. Considering their distinct numerical characteristics,
tailored preprocessing is introduced to improve the quantization efficiency.
Specifically, the distribution of the elements in $\mathbf{A}_{3}$
and $\mathbf{B}_{3}$ in the constellation diagram forms a uniform
ring pattern. This suggests that, after appropriate scaling and bias
adjustment, both the phase and amplitude of the two matrices exhibit
a uniform distribution. Since the center of $diag\left(\mathbf{A}_{3}\right)$
element distribution is offset from the origin, a bias adjustment
is necessary. On the other hand, while the center of $\mathbf{B}_{3}$
element distribution is at the origin, its amplitude range is relatively
large, requiring scaling before the quantization to ensure an optimal
representation. Consequently, for the diagonal matrix $\mathbf{A}_{3}$,
we subtract a bias before performing uniform quantization to its diagonal
elements. For $\mathbf{B}_{3}$, a scaling factor is applied before
the quantization to accommodate its broader value distribution. For
the real-valued vector $\mathbf{v}_{l_{m}}$, two quantization schemes
are employed: Uniform quantization and $\mu$-law quantization. 

Though the quantization bits of $\mathbf{v}_{l_{m}}$ basically determines
the overall feedback overhead, we find that the reconstruction error
exhibits greater sensitivity to $\Delta\mathbf{V}_{Y}$ which is solely
determined by the quantization error of $\mathbf{A}_{3}$ and $\mathbf{B}_{3}$.
Specifically, \eqref{eq:26} can be further expressed as \eqref{eq:36}
when considering the quantization error of all the feedback codewords.
The Hessian matrices of the objective function $f=\left\Vert \triangle\mathbf{H}_{s}\right\Vert _{F}^{2}$
with respect to $\Delta\mathbf{C}_{4}$ and $\Delta\mathbf{V}_{Y}$
are given in \eqref{eq:37} and \eqref{eq:38}, where $\hat{\mathbf{V}}_{Y}$
and $\hat{\mathbf{C}}_{4}$ denote the reconstruction matrices perturbed
by the quantization of codewords. Numerical results confirm that $\left\Vert \mathcal{H}_{\Delta\mathbf{V}_{Y}}f\right\Vert _{F}^{2}>\left\Vert \mathcal{H}_{\Delta\mathbf{C}_{4}}f\right\Vert _{F}^{2}$
indicating that the reconstruction error $\left\Vert \triangle\mathbf{H}_{s}\right\Vert _{F}^{2}$
is more sensitive to variations in $\Delta\mathbf{V}_{Y}$.\textcolor{blue}{{}
}This suggests that greater attention should be devoted to the quantization
bit allocation for $\mathbf{A}_{3}$ and $\mathbf{B}_{3}$. 
\begin{equation}
\left\Vert \triangle\mathbf{H}_{s}\right\Vert _{F}^{2}=\left\Vert \mathbf{H}_{s}-\left(\mathbf{C}_{4}+\Delta\mathbf{C}_{4}\right)\left(\mathbf{V}_{Y}+\Delta\mathbf{V}_{Y}\right)^{H}\right\Vert _{F}^{2}\label{eq:36}
\end{equation}
\begin{equation}
\begin{alignedat}{1}\mathcal{H}_{\Delta\mathbf{C}_{4}}f= & \left[\begin{array}{cc}
\left(\hat{\mathbf{V}}_{Y}^{H}\hat{\mathbf{V}}_{Y}\right)^{*}\otimes\mathbf{I}_{N_{t}} & \mathbf{0}\\
\mathbf{0} & \left(\hat{\mathbf{V}}_{Y}^{H}\hat{\mathbf{V}}_{Y}\right)\otimes\mathbf{I}_{N_{t}}
\end{array}\right]\\
\approx & \left[\begin{array}{cc}
\mathbf{I}_{r_{f}N_{t}} & \mathbf{0}\\
\mathbf{0} & \mathbf{I}_{r_{f}N_{t}}
\end{array}\right]
\end{alignedat}
\label{eq:37}
\end{equation}
\begin{equation}
\mathcal{H}_{\Delta\mathbf{V}_{Y}}f=\left[\begin{array}{cc}
\left(\hat{\mathbf{C}}_{4}^{H}\hat{\mathbf{C}}_{4}\right)^{*}\otimes\mathbf{I}_{2N} & \mathbf{0}\\
\mathbf{0} & \left(\hat{\mathbf{C}}_{4}^{H}\hat{\mathbf{C}}_{4}\right)\otimes\mathbf{I}_{2N}
\end{array}\right]\label{eq:38}
\end{equation}

Guided by this insight, we fine-tune the quantization bits of $\mathbf{A}_{3}$
and $\mathbf{B}_{3}$. Specifically, the UE detects potentially abnormal
reconstruction samples in advance before feedback by evaluating the
$\mathcal{F}$-norm of the gradient matrix. After discarding the second-order
terms, the gradient matrix of $\left\Vert \Delta\mathbf{H}_{s}\right\Vert _{F}^{2}$
can be expressed as \eqref{eq:39}.
\begin{equation}
\begin{split}\mathbf{G}=\frac{\partial\left\Vert \Delta\mathbf{H}_{s}\right\Vert _{F}^{2}}{\partial\Delta\mathbf{V}_{Y}}= & \left(\mathbf{C}_{4}^{H}\mathbf{C}_{4}\Delta\mathbf{V}_{Y}^{H}+\Delta\mathbf{C}_{4}^{H}\mathbf{C}_{4}\Delta\mathbf{V}_{Y}^{H}\right.\\
 & \left.\left(\mathbf{C}_{4}+\Delta\mathbf{C}_{4}\right)^{H}\Delta\mathbf{C}_{4}\left(\mathbf{V}_{Y}+\Delta\mathbf{V}_{Y}\right)^{H}\right)^{T}\\
\approx & \left(\mathbf{C}_{4}^{H}\mathbf{C}_{4}\Delta\mathbf{V}_{Y}^{H}+\mathbf{C}_{4}^{H}\Delta\mathbf{C}_{4}\mathbf{V}_{Y}^{H}\right)^{T}
\end{split}
\label{eq:39}
\end{equation}

Further, $\mathbf{G}$ can be written as $\mathbf{G}=\mathbf{G}_{1}+\mathbf{G}_{2}$
where:
\begin{equation}
\mathbf{G}_{1}=\left(\mathbf{C}_{4}^{H}\mathbf{C}_{4}\Delta\mathbf{V}_{Y}^{H}\right)^{T}\label{eq:40}
\end{equation}
\begin{equation}
\mathbf{G}_{2}=\left(\mathbf{C}_{4}^{H}\Delta\mathbf{C}_{4}\mathbf{V}_{Y}^{H}\right)^{T}\label{eq:41}
\end{equation}
Consequently, we can derive the upper bound for the $\mathcal{F}$-norm
of $\mathbf{G}$:
\begin{equation}
\left\Vert \mathbf{G}\right\Vert _{F}\leq\left\Vert \mathbf{G}_{1}\right\Vert _{F}+\left\Vert \mathbf{G}_{2}\right\Vert _{F}
\end{equation}

Since $\left\Vert \mathbf{G}_{2}\right\Vert _{F}$ is independent
of $\Delta\mathbf{V}_{Y}$, by focusing on $\left\Vert \mathbf{G}_{1}\right\Vert _{F}$
we can simplify the overall analysis and derive a useful insight into
how the CSI reconstruction error $\left\Vert \Delta\mathbf{H}_{s}\right\Vert _{F}^{2}$
behaves as a function of the quantization errors of the other two
matrices $\mathbf{A}_{3}$ and $\mathbf{B}_{3}$. Specifically, when
$\left\Vert \mathbf{G}_{1}\right\Vert _{F}$ has large values, it
indicates that the quantization system is far from the optimal state
and the reconstruction error is likely to be large. Note that $\left\Vert \mathbf{G}_{1}\right\Vert _{F}$
is fully known at the UE side, consequently providing an efficient
way to anticipate the potential recovery anomalies prior to feedback.
The proposed algorithm, as summarized in Algorithm\ \ref{alg:2},
can perform quantization adjustments on the CSI samples likely to
exhibit significant reconstruction errors, thereby enhancing the overall
robustness of the feedback scheme. 

\begin{algorithm}[tbh]
\caption{The proposed Robust Quantization Algorithm for $\mathbf{A}_{3}$ and
$\mathbf{B}_{3}$.\label{alg:2}}

\textbf{Input: }LI-MOR basis matrices: $\left\{ \mathbf{C}_{4},\mathbf{A}_{3},\mathbf{B}_{3}\right\} $;
Sampled frequency indices: $\left\{ f_{1},f_{2},...,f_{N}\right\} $;
Error threshold: $\epsilon$;

\textbf{Initialize:} Quantization bits (amplitude/phase) for $\left\{ \mathbf{A}_{3},\mathbf{B}_{3}\right\} $:
$\left(A,P\right)$ ; Adjustment step size for quantization bits:
$\left(\Delta A,\Delta P\right)$;

\textbf{Step1:} Obtain $\mathbf{Y}$ as \eqref{eq:22} with $\left\{ \mathbf{A}_{3},\mathbf{B}_{3}\right\} $
and $\left\{ f_{1},f_{2},...,f_{N}\right\} $; Perform thin SVD to
$\mathbf{Y}$ and obtain $\mathbf{V}_{Y}$ following \eqref{eq:24};

\textbf{Step2:} Perform uniform quantization to $\left\{ \mathbf{A}_{3},\mathbf{B}_{3}\right\} $
with $\left(A,P\right)$ bits and reconstruct $\left\{ \hat{\mathbf{A}}_{3},\hat{\mathbf{B}}_{3}\right\} $;
Compute $\mathbf{\hat{V}}_{Y}$ with $\left\{ \hat{\mathbf{A}}_{3},\hat{\mathbf{B}}_{3}\right\} $
as \eqref{eq:22} and \eqref{eq:24}; Obtain the error matrix $\Delta\mathbf{V}_{Y}=\mathbf{\hat{V}}_{Y}-\mathbf{V}_{Y}$;
Obtain $\mathbf{G}_{1}$ as \eqref{eq:40}; 

\textbf{Step3:} If $\left\Vert \mathbf{G}_{1}\right\Vert _{F}\geq\epsilon$,
$A=A+\Delta A,P=P+\Delta P$ and go to \textbf{Step 2};

\textbf{Step4:} If $\left\Vert \mathbf{G}_{1}\right\Vert _{F}<\epsilon$,
terminate the algorithm;

\textbf{Output:} Binary codes of quantized $\left\{ \mathbf{A}_{3},\mathbf{B}_{3}\right\} $
with $\left(A,P\right)$: $\mathbf{b}_{\mathbf{A}},\mathbf{b}_{\mathbf{B}}$; 
\end{algorithm}

\subsection{Summary of the Overall Framework}

\begin{figure*}[t]
\begin{centering}
\includegraphics[width=16.5cm,totalheight=5.5cm]{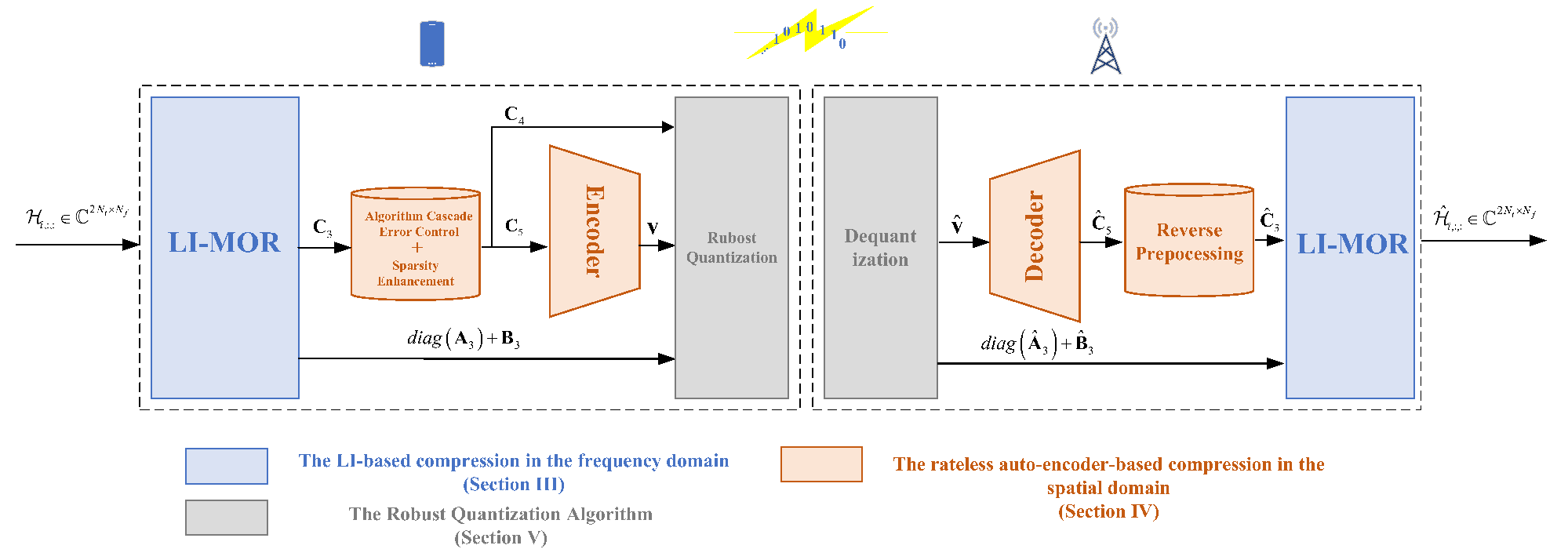}
\par\end{centering}
\caption{The workflow of the overall framework.\label{fig:4}}

\end{figure*}

The overall workflow of the proposed wideband CSI feedback scheme,
which integrates the LI framework, DL-based spatial compression and
the robust quantization module, is illustrated in Fig.\ \ref{fig:4}.

At the UE side, the original channel slice $\mathbf{H}=\mathcal{H}_{i,:,:}\in\mathbb{C}^{2N_{t}\times N_{f}}$
corresponding to the $i$-th receive antenna is initially compressed
in the frequency domain via LI-MOR (the blue block in Fig.\ \ref{fig:4}
as introduced in Section\ \ref{sec:3}). This process constructs
the LI bases $\left\{ \mathbf{C}_{3},\mathbf{B}_{3},\mathbf{A}_{3}\right\} $.
Among them, $\mathbf{B}_{3}$ and $\mathbf{A}_{3}$ exhibit simple
structures, whereas $\mathbf{C}_{3}\in\mathbb{C}^{N_{t}\times r_{f}}$
still retains redundant information in the spatial domain, especially
when $N_{t}$ is large for massive MIMO systems. Subsequently, $\mathbf{C}_{3}$
undergoes preprocessing through unitary matrix merging and sparsity
enhancement to control the cascaded errors before being fed into the
rateless auto-encoder (the orange block in Fig.\ \ref{fig:4} as
introduced in Section\ \ref{sec:4}). After the compression in both
the frequency and spatial domains, the resulting codewords are given
by $\mathbf{z}_{i}=\left\{ \mathbf{v}_{t},\mathbf{A}_{3},\mathbf{B}_{3}\right\} $,
to which different quantization strategies are applied to ensure the
robustness of the overall scheme across different channel samples
(the gray block in Fig.\ \ref{fig:4} as introduced in Subsection\ \ref{subsec:5-1}). 

At the BS side, the received binary bits are first dequantized according
to the corresponding quantization scheme, yielding $\mathbf{\hat{z}}_{i}=\left\{ \hat{\mathbf{v}}_{t},\hat{\mathbf{A}}_{3},\hat{\mathbf{B}}_{3}\right\} $.
The codeword $\hat{\mathbf{v}}_{t}$ is fed into the decoder to reconstruct
$\hat{\mathbf{C}}_{5}$. By applying the reverse preprocessing described
as \eqref{eq:29}, the BS further reconstructs $\hat{\mathbf{C}}_{3}$.
With $\left\{ \hat{\mathbf{C}}_{3},\hat{\mathbf{A}}_{3},\hat{\mathbf{B}}_{3}\right\} $,
the BS can recover $\hat{\mathbf{H}}$ via the interpolation function
defined in\textcolor{blue}{{} }\eqref{eq:17}. It is worth noting that
the proposed variable-length feedback operates at the codeword level.
Therefore, the UE needs to append several additional bits to indicate
the quantization bit width of each codeword (the overhead of these
indicator bits is negligible), enabling proper dequantization at the
BS. 

\section{Simulation Results\label{sec:6}}

In this section, we evaluate the performance of the proposed wideband
CSI feedback scheme. First, we describe the parameter settings of
the wideband channels and the dataset. Then, we present the simulation
results including the feedback precision and overhead, the computational
complexity, the parameter storage and the spectral efficiency of the
proposed algorithm and the baseline schemes. 

\subsection{Parameter Setting and Channel Data}

We consider the wideband CDL-A and CDL-B channels in 3GPP 38.901\ \cite{3gpp901}.
To facilitate the simulation comparison, we separate the $400M$ wideband
channel matrix and take one of the $100M$ bandwidth as an example.
The main parameters are listed in Table\ \ref{tab:1}. Note that
the size of the feedback channel slice for the proposed scheme is
$\mathcal{H}_{i,:,:}\in\mathbb{C}^{256\times3300}$. When generating
the channel dataset, different random seeds (1\textasciitilde 100000)
are used to ensure that the resulting channel samples are non-overlapping.
The dataset is partitioned into 50,000 samples for training, 10,000
for validation, and 10,000 for testing.

\begin{table}[tbh]
\caption{{\footnotesize\label{tab:1}MAIN CHANNEL PARAMETERS}}

\centering{}%
\begin{tabular}{|c|c|}
\hline 
Channel profile & CDL-A/CDL-B (NLOS)\tabularnewline
\hline 
Carrier center frequency $f_{c}$ & $6.9$ GHz\tabularnewline
\hline 
Transmitting antenna $N_{t}$ & $16\times8\left(UPA\right)$\tabularnewline
\hline 
Receiving antenna $N_{r}$ & 2\tabularnewline
\hline 
Antenna polarization & $\pm45^{\lyxmathsym{\textdegree}}$\tabularnewline
\hline 
RMS Delay spread & $30$ ns\tabularnewline
\hline 
Number of resource blocks$N_{RB}$ & $275$\tabularnewline
\hline 
Subcarrier spacing & $30$ kHz\tabularnewline
\hline 
Number of Subcarriers $N_{f}$ & $12\times275=3300$\tabularnewline
\hline 
Bandwidth & $100$ MHz\tabularnewline
\hline 
Number of layer $R$ & 2\tabularnewline
\hline 
\end{tabular}
\end{table}

\subsection{NMSE and Feedback Overhead Performance}

The normalized MSE (NMSE) is used to evaluate the recovery accuracy
of CSI matrix $\mathbf{H}=\mathcal{H}_{i,:,:}\in\mathbb{C}^{256\times3300}$,
as defined in \eqref{eq:43}. Our overall principle is to make a fair
comparison by showing that the proposed scheme can achieve a smaller
reconstruction error with equal or less feedback overhead which is
either the complex numbers (when evaluating the performance without
quantization) or the number of the binary bits (when evaluating the
performance with quantization).
\begin{equation}
NMSE=E\left\{ \left\Vert \widehat{\mathbf{H}}-\mathbf{H}\right\Vert _{F}^{2}/\left\Vert \mathbf{H}\right\Vert _{F}^{2}\right\} \label{eq:43}
\end{equation}

\begin{figure}[tbh]
\begin{centering}
\includegraphics[width=8cm,totalheight=4cm]{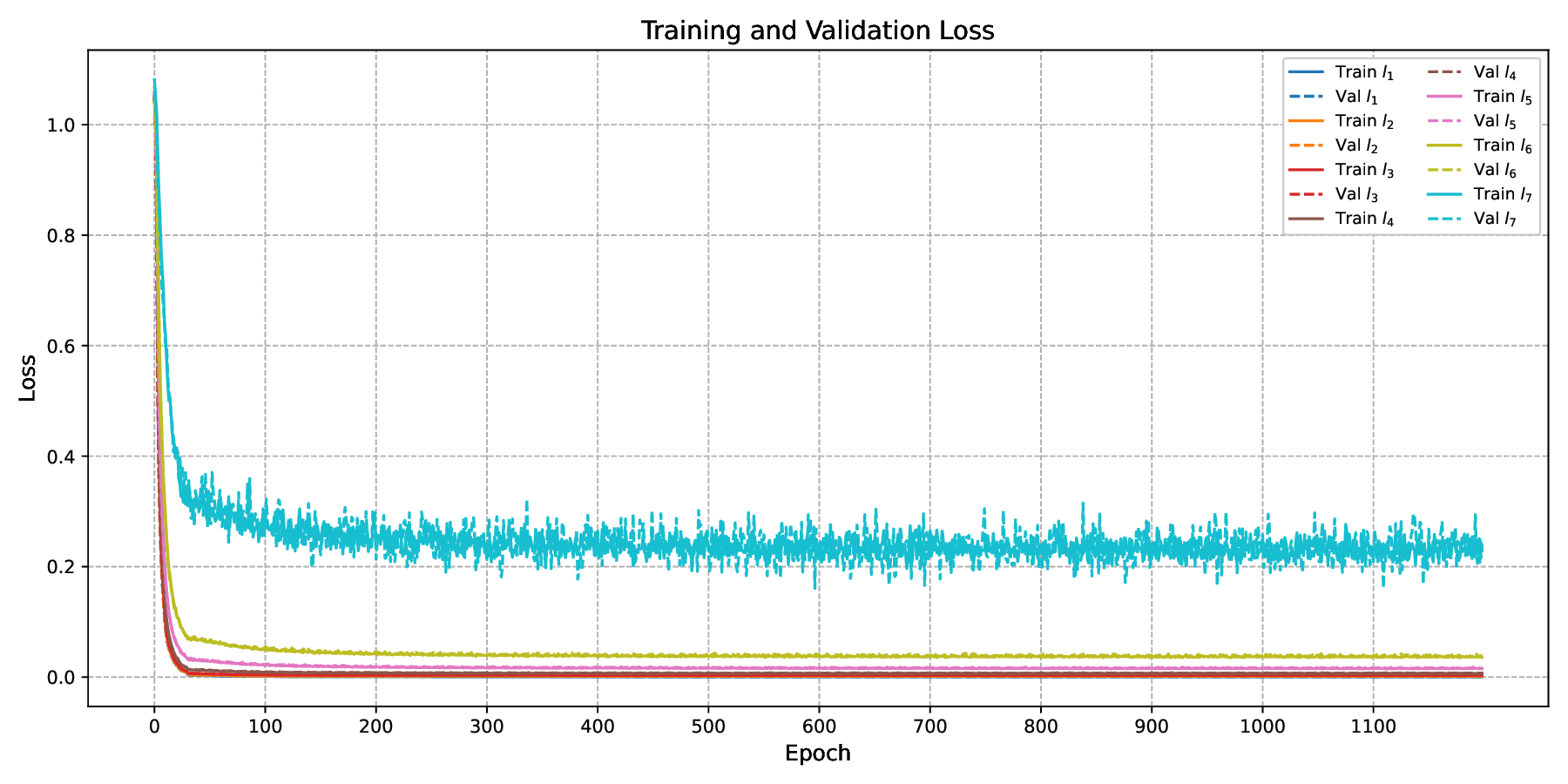}
\par\end{centering}
\caption{The Training Curve of the LI-MORNet.\label{fig:5}}
\end{figure}

We compare the proposed LI-MORNet to the following baseline schemes. 
\begin{itemize}
\item \textbf{SM-CsiNet+\ \cite{DL_5}:} The encoder achieves multi-rate
feedback through multiple cascaded FC layers. The BS must deploy multiple
decoders corresponding to the CRs. To extend the set of compression-rate
nodes, the encoder architecture must be modified, additional decoders
must be deployed at the BS, and the entire model must be retrained.
\item \textbf{PM-CsiNet+\ \cite{DL_5}:} The feature extraction module
is the same as that of SM-CsiNet+. But the encoder achieves multi-rate
feedback through parallel FC layers which is a more compact model.
The BS still needs multiple decoders. 
\item \textbf{CL-Net\ \cite{DL_3}:} A high-precision CSI feedback network
that effectively integrates the real and imaginary parts and utilizes
spatial attention mechanism. However, each pre-trained model supports
feedback at only one CR. Changing the CR requires retraining the model
and deploying additional encoder-decoder pairs at both the UE and
the BS. 
\end{itemize}
All the baseline schemes employ the DFT for channel characterization
as the preprocessing, which involves reconstructing the sparse angular-delay
domain channel representation $\mathbf{\tilde{H}}=\mathbf{F}_{d}\mathbf{H}\mathbf{F}_{a}^{H}$
. Following the truncation ratio adopted in prior work\ \cite{DL_1},
the number of retained columns in the delay domain is set to $Nc=103$.
Truncation is applied only in the delay domain because it has substantially
higher dimension and exhibits stronger sparsity compared with the
angular domain. All networks were trained for 1,000 epochs, using
a cosine annealing schedule for the learning rate. For the proposed
LI-MORNet, the initial codeword length $M$ is set to $N_{t}r_{f}$
so that the initial $CR_{s}$ is $1/2$ and the interval length of
the random mask is set to $\Delta L=255$ with $n=7,L_{1}^{min}=256,L_{7}^{max}=2048$
and $L_{i}^{max}+1=L_{i+1}^{min},\forall i=1,...,n-1$ so that the
supported range of the CR by the rateless auto-encoder is $\frac{1}{32}\leq CR_{s}\leq\frac{1}{4}$.
The weight coefficients in \eqref{eq:34} is set to $\left\{ w_{1},w_{2},w_{3},w_{4},w_{5},w_{6},w_{7}\right\} =\left\{ 25,20,10,5,1,1,1\right\} $.
The training curve of LI-MORNet is shown in Fig.\ \ref{fig:5}.

\subsubsection{Performance without Quantization}

\begin{figure}[tbh]
\begin{centering}
\subfloat[CDL-A]{\begin{centering}
\includegraphics[width=7.8cm,totalheight=5.7cm]{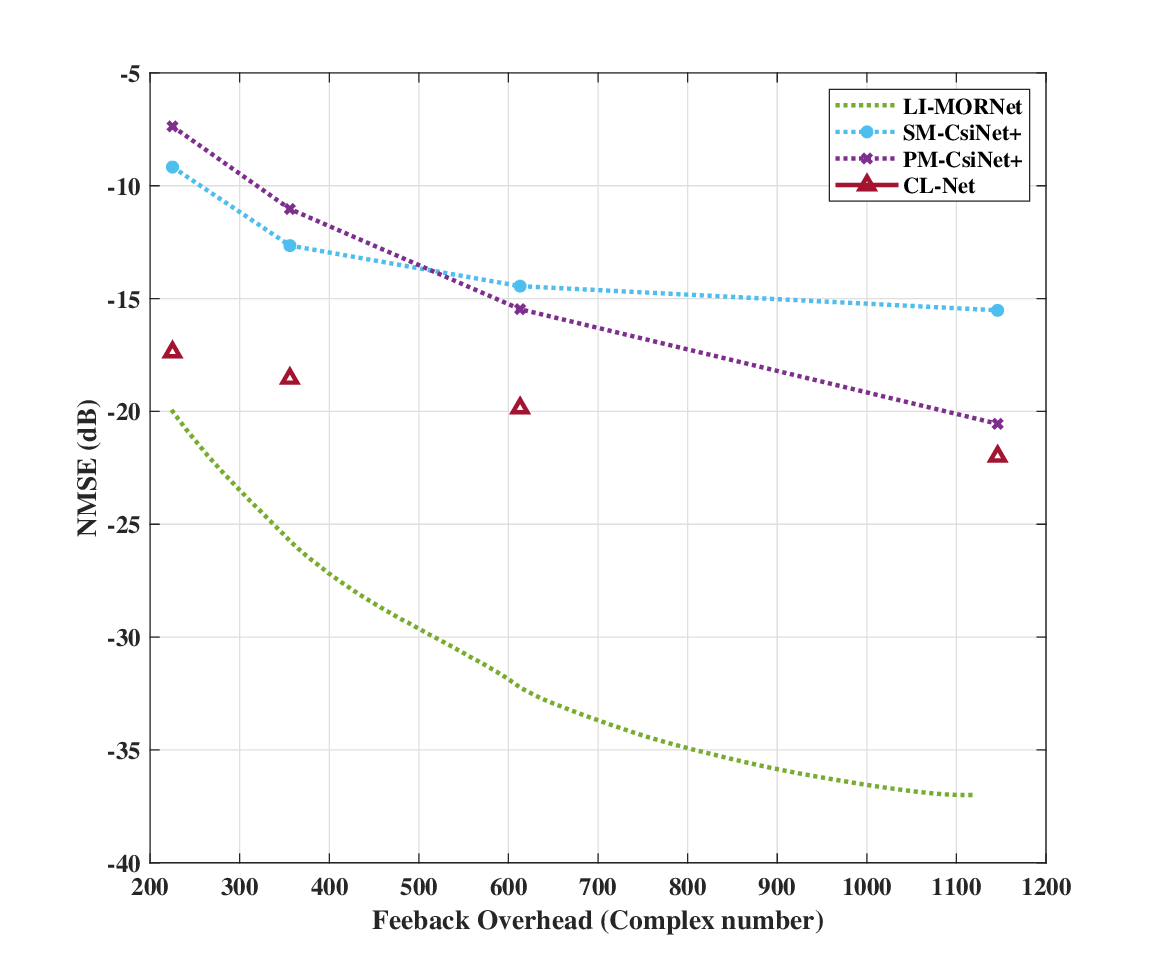}
\par\end{centering}

}
\par\end{centering}
\begin{centering}
\subfloat[CDL-B]{\begin{centering}
\includegraphics[width=7.8cm,totalheight=5.7cm]{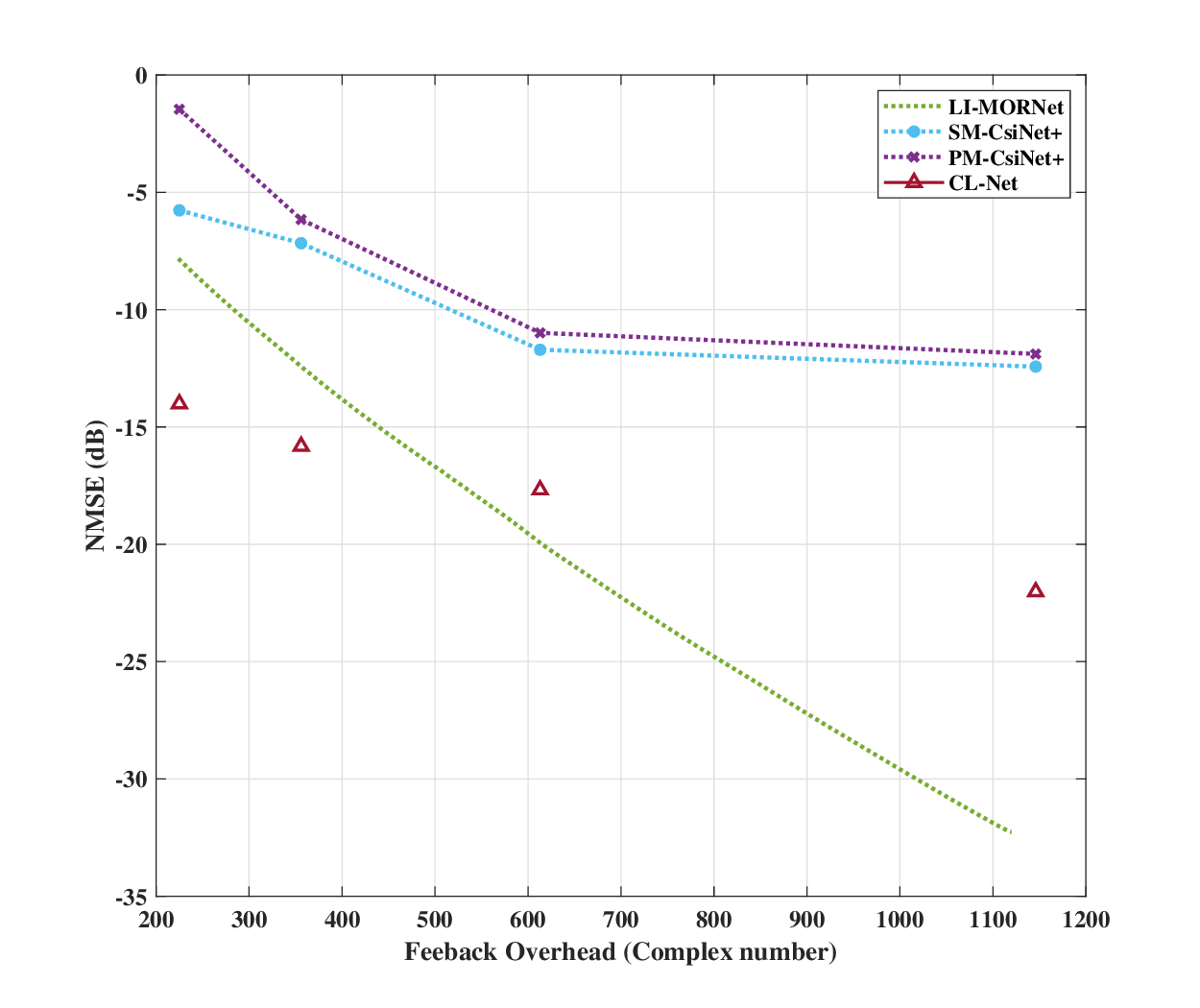}
\par\end{centering}

}
\par\end{centering}
\caption{The NMSE performance without quantization of different CSI feedback
schemes.\label{fig:6}}

\end{figure}

Fig.\ \ref{fig:6} illustrates the performance comparison of different
CSI feedback schemes without quantization. For the two multi-rate
feedback algorithms, the pre-trained model supports only the CRs corresponding
to the nodes on the curve. For the single-rate feedback algorithm,
each node represents an individual model pre-trained for a specific
CR. In contrast, our proposed LI-MORNet enables a single model to
support feedback at any point along the curve. Based on the simulation
results, our proposed model achieves much higher reconstruction precision
than all multi-rate feedback schemes, and it even outperforms the
single-rate network (CL-Net) for most cases. It only exhibits lower
reconstruction accuracy than the single-rate network CL-Net under
low feedback overhead for the CDL-B channel. However, even for this
case, it still offers clear advantages in smaller parameter storage
and computational complexity and the flexibility of the trade-off
between feedback overhead and reconstruction accuracy, as will be
further analyzed in Subsection\ \ref{subsec:6-3}.

The number of feedback complex numbers of our proposed LI-MORNet is
$d=\left(N_{t}\times r_{f}\times CR_{s}+r_{f}\times3\right)$ where
$CR_{s}$ is defined as \eqref{eq:19}. We make an initial selection
of $r_{f}$ according to the number of paths in the delay domain and
then fine-tune it according to the numerical simulation results. With
$r_{f}=32$ and $\frac{1}{32}\leq CR_{s}\leq\frac{1}{4}$, the overall
CR which is defined as $CR=\nicefrac{d}{2N_{t}N_{sc}}$ that we compare
falls in the range $\frac{1}{3770}\leq CR\leq\frac{1}{800}$.

\subsubsection{Impact of the Robust Quantization Algorithm}

\begin{figure}[tbh]
\begin{raggedright}
\subfloat[CDL-A]{\begin{centering}
\includegraphics[width=9.2cm,totalheight=3.8cm]{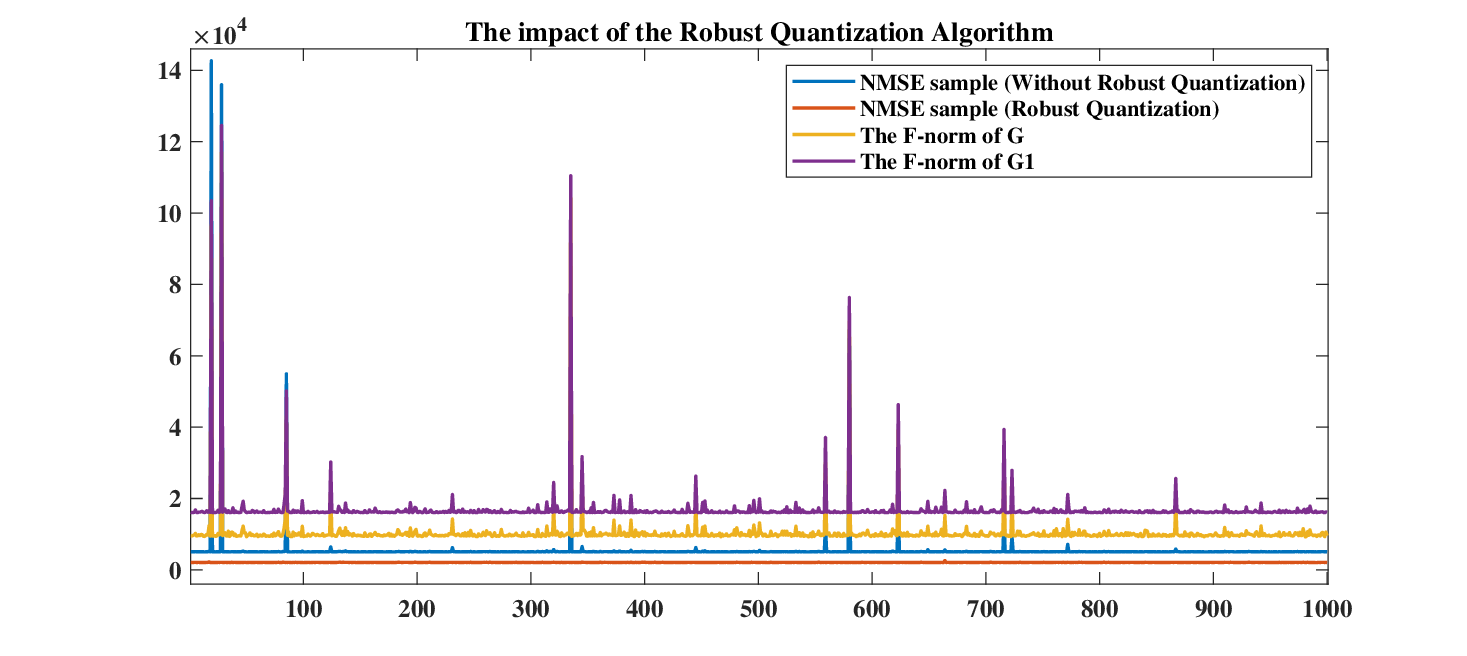}
\par\end{centering}
}
\par\end{raggedright}
\begin{raggedright}
\subfloat[CDL-B]{\begin{centering}
\includegraphics[width=9.2cm,totalheight=3.8cm]{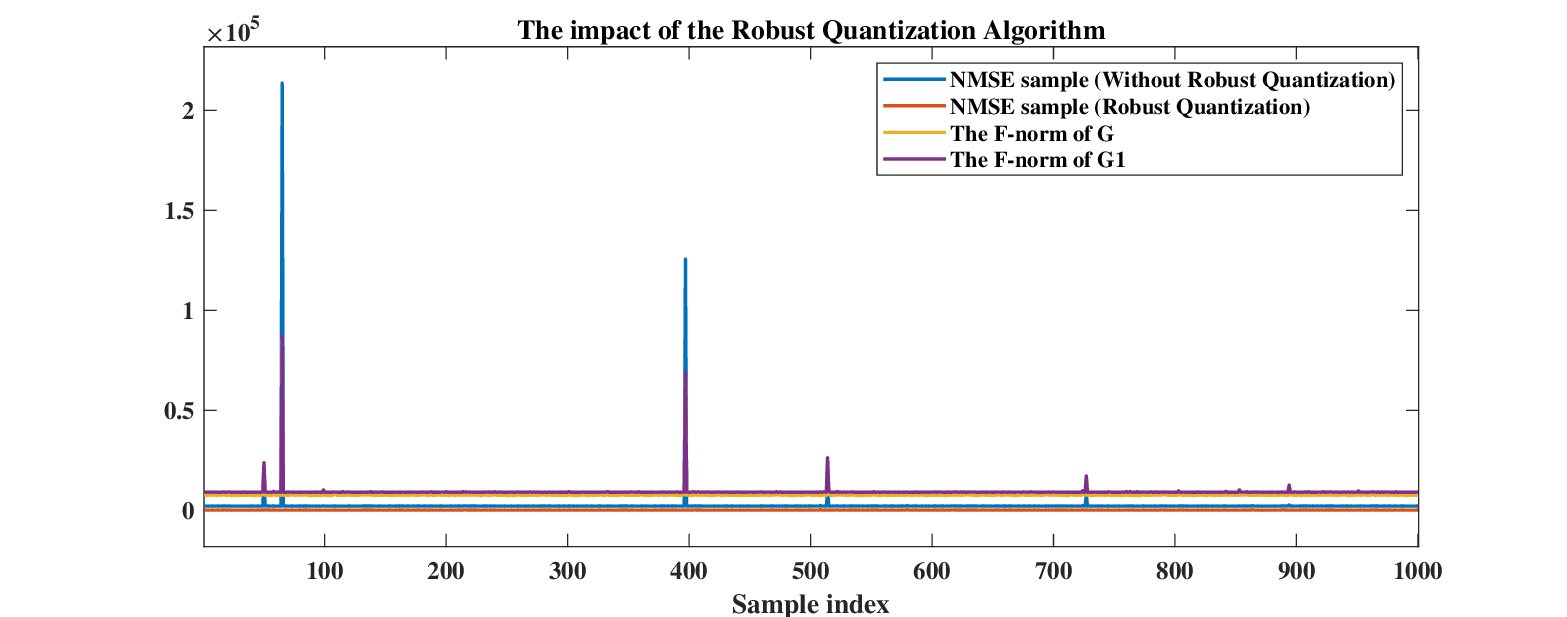}
\par\end{centering}
}
\par\end{raggedright}
\caption{The impact of the Robust Quantization Algorithm.\label{fig:7}}
\end{figure}

Fig.\ \ref{fig:7} illustrates the effectiveness of the proposed
robust quantization scheme under CDL-A and CDL-B channel models. The
horizontal axis denotes the channel sample index. For clarity of presentation,
1000 random samples are selected from the test set and shown in the
figure. Note that to distinguish between the curves, we added different
offsets when plotting the graph. 

As introduced in Subsection\ \ref{subsec:5-1}, the yellow curve
represents $\left\Vert \mathbf{G}\right\Vert _{F}$, which quantifies
the sensitivity of the quantization error of $\mathbf{A}_{3},\mathbf{B}_{3}$
with respect to the perturbation at the sampling point. The purple
curve denotes $\left\Vert \mathbf{G}_{1}\right\Vert _{F}$, a constant
known at the UE that serves as an indicator for identifying samples
potentially prone to unstable or anomalous recovery. It can be observed
from the results that both the purple and yellow curves remain relatively
small for the majority of channel samples. However, at several sample
indices, the curves exhibit pronounced peaks. These peaks indicate
that the system is momentarily far from its optimal point, leading
to a larger reconstruction error at the BS. This sensitivity behavior
is directly confirmed by the blue curve, which plots the NMSE using
the conventional quantization strategy (without the robust quantization
scheme). The peaks nearly coincide with those of the yellow detection
curve which means that with the known $\left\Vert \mathbf{G}_{1}\right\Vert _{F}$
at the UE side the potential low-accuracy recovery samples can be
accurately detected. The orange curve further demonstrates that, with
the proposed robust quantization scheme, these anomalous samples can
be reconstructed with high precision, and the recovery at the BS remains
stable across the entire sample set.

\subsubsection{Performance with Quantization}

\begin{figure}[tbh]
\begin{centering}
\subfloat[CDL-A]{\begin{centering}
\includegraphics[width=7.8cm,totalheight=5.7cm]{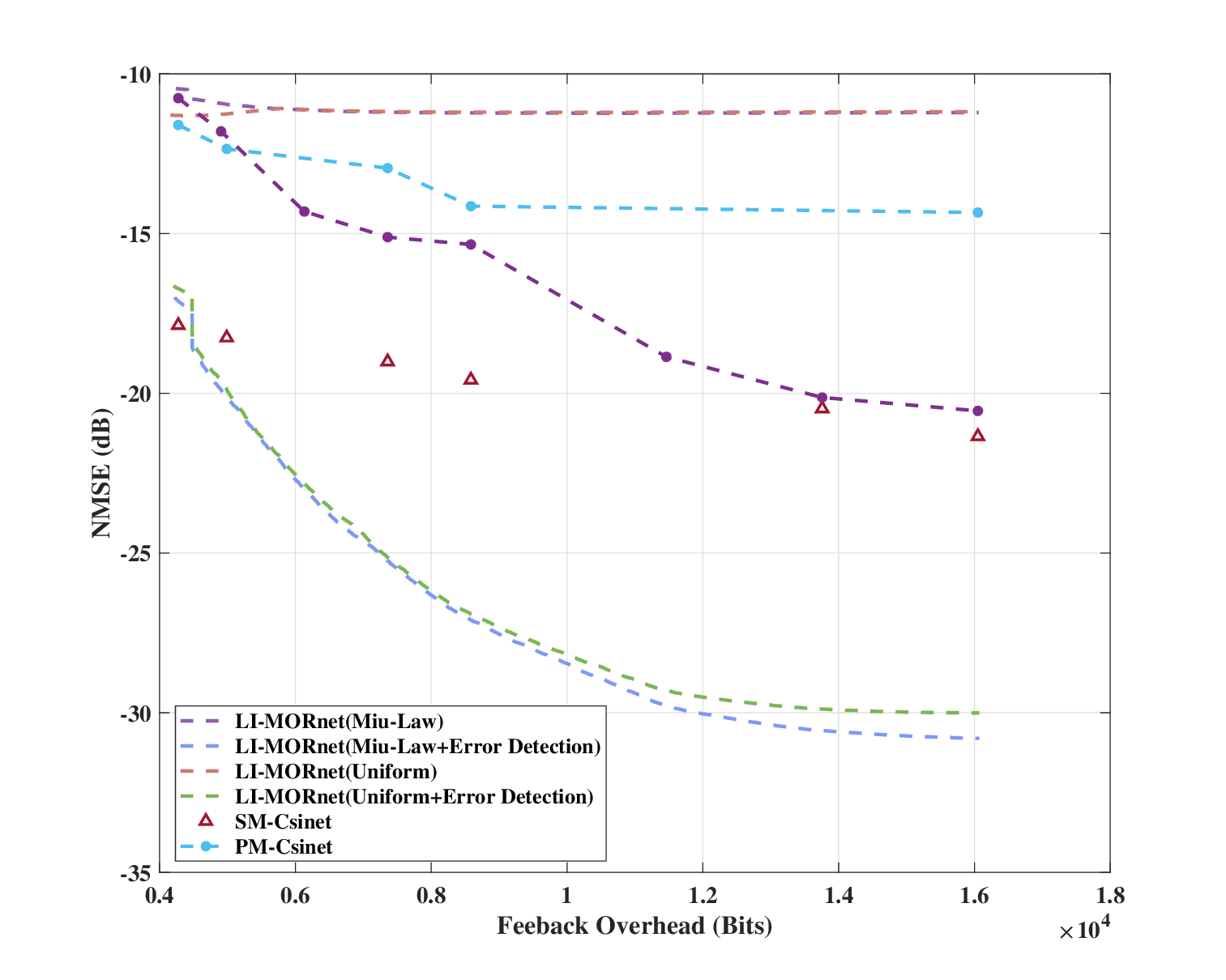}
\par\end{centering}

}
\par\end{centering}
\begin{centering}
\subfloat[CDL-B]{\begin{centering}
\includegraphics[width=7.8cm,totalheight=5.7cm]{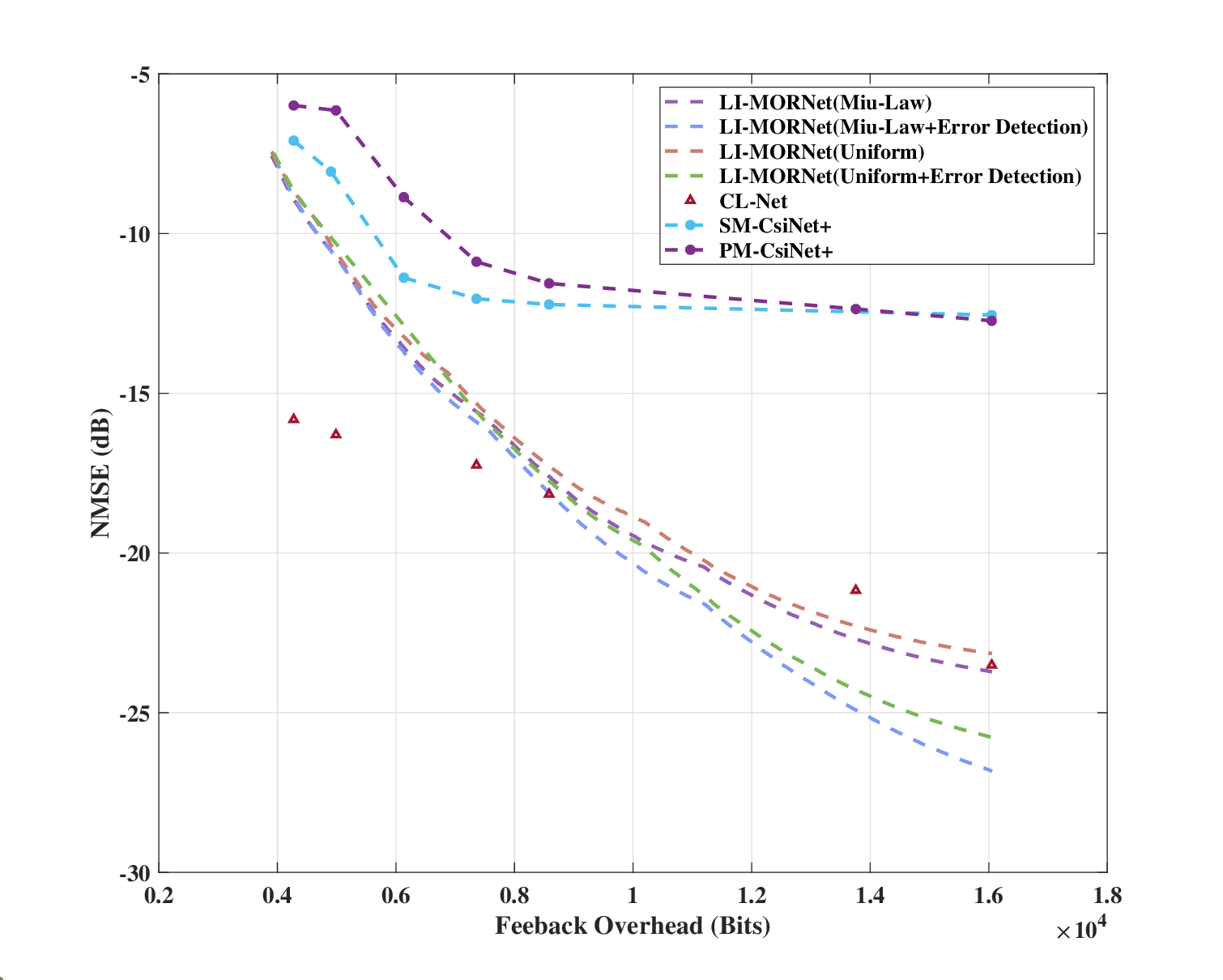}
\par\end{centering}

}
\par\end{centering}
\caption{The NMSE performance with quantization of different CSI feedback schemes.\label{fig:8}}

\end{figure}

Fig.\ \ref{fig:8} shows the performance comparison of different
CSI feedback schemes with quantization. For all baseline algorithms,
we applied both uniform quantization and \textgreek{\textmu}-law quantization
using codeword lengths of $4\sim7$ bits. We then select the optimal
configuration by jointly considering the compression ratio, quantization
scheme (i.e., number of bits and quantization method), and feedback
accuracy. Specifically, among the configurations with comparable feedback
overhead, we chose the one that achieves the highest feedback accuracy.

\subsection{Computational Complexity Comparison\label{subsec:6-3}}

We use the number of floating point operations (FLOPs) to evaluate
the NN complexity. The total computational complexity of the proposed
LI-MORNet consists of two major components.

For the LI-based compression at the UE, only the dominant term is
considered in the final result, as expressed in \eqref{eq:12}. For
the SVD of a $m\times n$ matrix, retaining the leading $k$ singular
components via rank-revealing QR factorization requires $\mathcal{O}\left(mnk\right)$
FLOPs~{[}11{]}{[}12{]}. Consequently, the computational complexity
of the compression in the frequency domain at the UE is $\mathcal{O}\left(2qpN_{T}r_{f}\right)$.
At the BS side, the dominant computational complexity at the BS arises
from the interpolation over the full set of subcarriers as shown in
\eqref{eq:17}. Since $\mathbf{A}_{3}$ is a diagonal matrix and $\mathbf{B}_{3}\in\mathbb{C}^{r_{f}\times2}$
, we calculate $\widehat{\mathbf{H}}\left(f_{p}\right)=\hat{\mathbf{C}}_{3}\left(f_{p}\mathbf{I}-\mathbf{\hat{A}}_{3}\right)^{-1}\mathbf{\hat{B}}_{3}\in\mathbb{C}^{N_{T}\times2}$
for all $N_{sc}$ subcarriers, yielding a total computational complexity
of $\mathcal{O}\left(2N_{T}N_{sc}r_{f}\right)$. Therefore, the overall
algorithmic complexity of the LI-based compression and decompression
is $\mathcal{O}\left(2qpN_{T}r_{f}\right)+\mathcal{O}\left(2N_{T}N_{sc}r_{f}\right)$.
The typical values of $p$ and $q$ are $137$ and $138$ respectively.

For the compression and the decompression in the spatial domain, the
FLOPs of the rateless auto-encoder are 91.767M. We combine the computational
complexity in both the frequency and the spatial domains for our proposed
scheme to ensure a fair comparison with the baselines. For the single-rate
baseline network, we include the parameter storage and FLOPs of all
supported CR branches.

Table\ \ref{tab:2} presents the comparison of the computational
complexity and parameter storage of different CSI feedback schemes.
Compared to the baseline networks, our proposed scheme achieves reductions
in both parameter storage and total computational complexity. In fact,
our scheme offers lower neural-network complexity, while the LI-based
compression in the frequency domain introduces additional computational
complexity, yet enables higher reconstruction accuracy and reduced
parameter storage. 

\begin{table}[tbh]
\caption{Complexity and Storage Comparison\label{tab:2}}

\centering{}%
\begin{tabular}{|c|c|c|c|c|}
\hline 
Scheme & LI-MORNet & CL-Net & SM-CsiNet+ & PM-CsiNet+\tabularnewline
\hline 
\hline 
FLOPs & 428.05M & 498.2M & 788.1M & 786.4M\tabularnewline
\hline 
Storage & 67.1M & 123.5M & 93.1M & 91.3M\tabularnewline
\hline 
\end{tabular}
\end{table}

\subsection{Spectral Efficiency Comparison}

\begin{figure}[tbh]
\begin{centering}
\subfloat[CDL-A]{\begin{centering}
\includegraphics[width=7.8cm,totalheight=5.7cm]{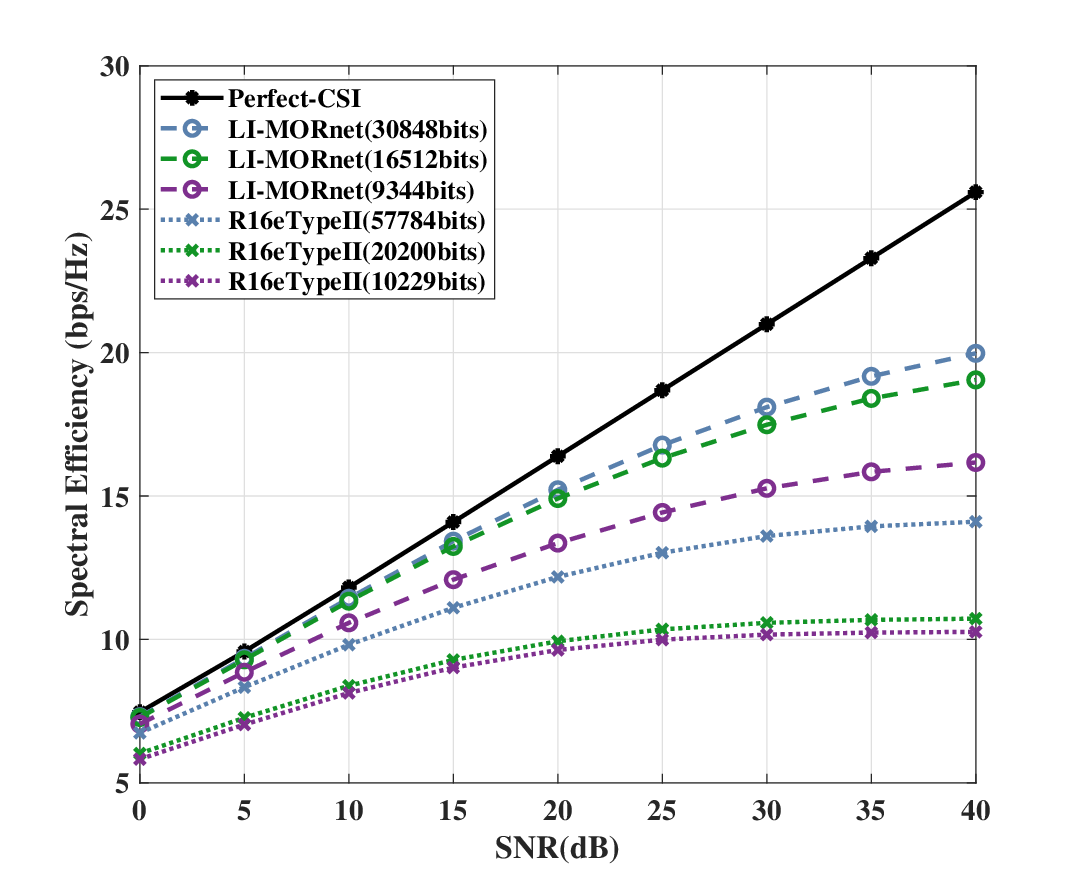}
\par\end{centering}

}
\par\end{centering}
\begin{centering}
\subfloat[CDL-B]{\begin{centering}
\includegraphics[width=7.8cm,totalheight=5.7cm]{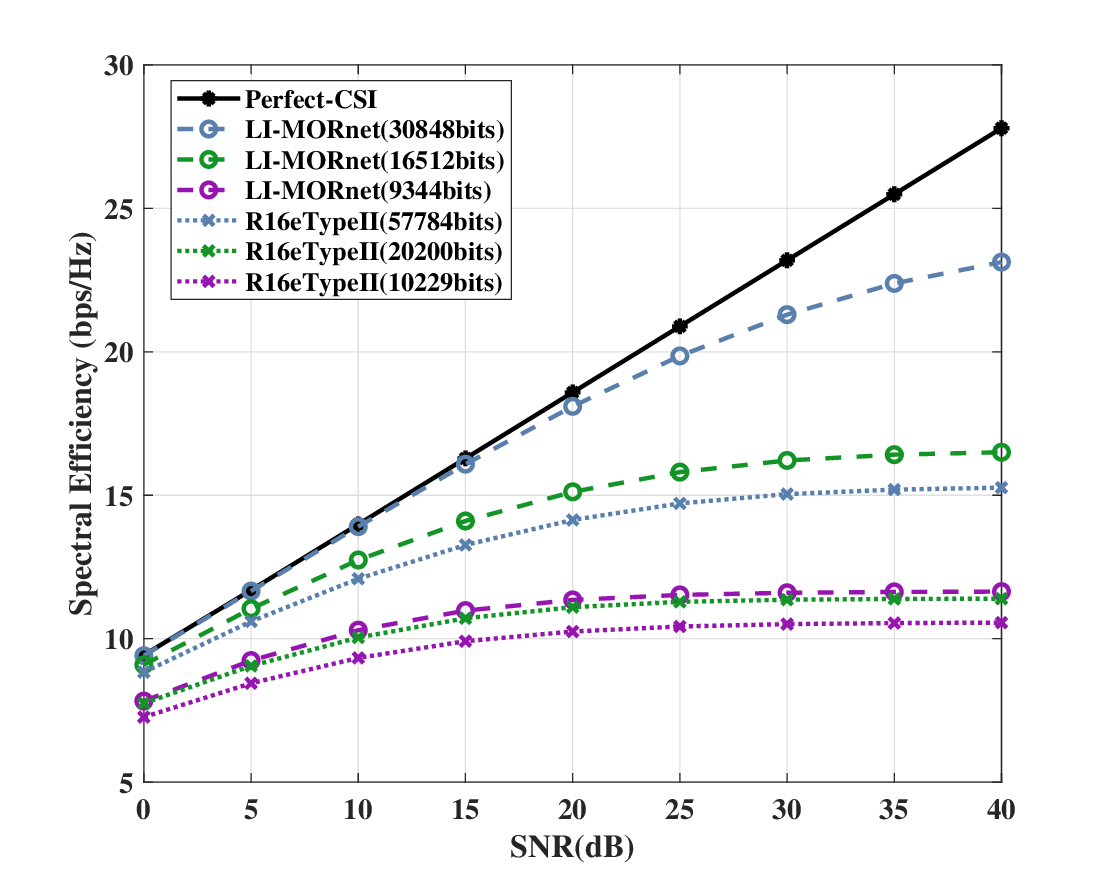}
\par\end{centering}

}
\par\end{centering}
\caption{The performance of the spectral efficiency.\label{fig:9}}

\end{figure}

In this section, we adopt the R16 eType-II codebook (mode 2,4,6) as
the baseline for spectral-efficiency comparison because eType-II and
our proposed scheme feed back different types of CSI respectively
(precoder-related information $\mathbf{W}$ versus the full channel
matrix $\mathcal{H}$). As a result, directly comparing their NMSE
would be neither fair nor physically meaningful. Instead, we evaluate
both schemes under the same downlink precoding framework and compare
their achievable spectral efficiency, thereby providing a consistent
and meaningful assessment of end-to-end system performance. The simulation
results are illustrated in Fig.\ \ref{fig:9}. In the considered
setup, the UE first obtains a perfectly estimated downlink channel
$\mathcal{H}=\left[\mathbf{H}_{1},\mathbf{H}_{2},...,\mathbf{H}_{N_{f}}\right]\in\mathbb{C}^{N_{r}\times2N_{t}\times N_{f}}$
and derives the receive combining vectors from its singular value
decomposition (SVD). Meanwhile, the BS reconstructs the channel information
based on the feedback and applies the corresponding SVD-based precoding
to facilitate downlink transmission as described in Section\ \ref{sec:2}.
Based on this configuration, the achievable spectral efficiency can
be expressed as $\frac{1}{N_{f}}\sum_{f=1}^{N_{f}}\sum_{l=1}^{R}\Gamma_{l,f}$
where 
\[
\Gamma_{l,f}=log\left(1+\frac{\left|\mathbf{u}_{l,f}^{H}\mathbf{H}_{f}\hat{\mathbf{v}}_{l,f}\right|^{2}}{\sum_{j=1,j\neq l}^{R}\left|\mathbf{u}_{j,f}^{H}\mathbf{H}_{f}\hat{\mathbf{v}}_{j,f}\right|^{2}+\frac{P_{n}}{P_{t}}\left\Vert \mathbf{u}_{l,f}\right\Vert ^{2}}\right)
\]
$\mathbf{H}_{f}\in\mathbb{C}^{N_{r}\times2N_{t}}$ represents the
channel matrix at the $f$-th subband. $\mathbf{u}_{l,f}$ denotes
the combining vector for stream $l$ of subband $f$, obtained using
the perfectly estimated channel matrix $\mathbf{H}_{f}=\mathbf{U}_{f}\mathbf{\varSigma}_{f}\mathbf{V}_{f}^{H}$.
Specifically, the $l$-th column of $\mathbf{U}_{f}$, denoted as
$\mathbf{u}_{l,f}$, is adopted as the receive combiner at the UE.
While the right singular vector $\mathbf{\hat{v}}_{l,f}$ of the reconstructed
channel matrix $\hat{\mathbf{H}}_{f}=\hat{\mathbf{U}}_{f}\mathbf{\hat{\varSigma}}_{f}\hat{\mathbf{V}}_{f}^{H}$
(the $l$-th column of $\hat{\mathbf{V}}_{f}$) is used as the corresponding
transmit precoder at the BS. This setup reflects a practical FDD feedback
scenario, where the UE utilizes the perfectly estimated channel to
compute the receive combining vectors, while the BS reconstructs the
CSI from feedback and applies the corresponding SVD-based precoding
for downlink transmission.

\section{Conclusion\label{sec:Conclusion}}

We propose a variable-length wideband CSI feedback scheme based on
LI and DL, enabling high-precision CSI reconstruction while offering
a flexible trade-off between feedback overhead and reconstruction
accuracy within the quantization framework. In the frequency domain,
we develop an LI-based CSI interpolation model that constructs a set
of initial bases, ensuring perfect CSI recovery at sampled subcarriers
and high-precision fitting across the remaining frequency indices.
With the MOR in the LI framework and the further simplification ,
two of the bases become structurally sparse and lightweight. This
allows us to focus on compressing the dominant basis, which still
contains spatial redundancy and has the dimension determined by the
number of antennas. We design a cascaded spatial compression scheme
that employs matrix transformations to prevent the cascaded error
amplification and enhance sparsity. Variable-length codeword feedback
is then enabled through a rateless auto-encoder. In addition, we develop
dedicated quantization strategies for the codewords derived from both
the frequency-domain and spatial-domain compression, improving the
robustness of the overall system across different channel samples.
The simulation results demonstrate that our scheme significantly outperforms
the baseline schemes in both CSI recovery accuracy and the communication
system performance on wideband channels. 

\bibliographystyle{unsrt}
\bibliography{refernece/reference_LI_CNN}

\end{document}